\documentclass[journal]{IEEEtran}
\usepackage[T1]{fontenc}% optional T1 font encoding
\usepackage{cite}
\usepackage{amsmath,amssymb,amsfonts}
\usepackage{algorithmic}
\usepackage[ruled,linesnumbered]{algorithm2e}
\usepackage{graphicx}
\usepackage{textcomp}
\usepackage{xcolor}
\usepackage{subfigure}
\usepackage{booktabs}
\usepackage{multirow}
\usepackage{diagbox}
\usepackage{url}
\usepackage{threeparttable}
\usepackage{array}
\usepackage{etoolbox}
\makeatletter
\def\UrlAlphabet{%
      \do\a\do\b\do\c\do\d\do\e\do\f\do\g\do\h\do\i\do\j%
      \do\k\do\l\do\m\do\n\do\o\do\p\do\q\do\r\do\s\do\t%
      \do\u\do\v\do\w\do\x\do\y\do\z\do\A\do\B\do\C\do\D%
      \do\E\do\F\do\G\do\H\do\I\do\J\do\K\do\L\do\M\do\N%
      \do\O\do\P\do\Q\do\R\do\S\do\T\do\U\do\V\do\W\do\X%
      \do\Y\do\Z}
\def\UrlDigits{\do\1\do\2\do\3\do\4\do\5\do\6\do\7\do\8\do\9\do\0}
\g@addto@macro{\UrlBreaks}{\UrlOrds}
\g@addto@macro{\UrlBreaks}{\UrlAlphabet}
\g@addto@macro{\UrlBreaks}{\UrlDigits}
\patchcmd{\@algocf@start}% <cmd>
  {-1.5em}% <search>
  {0pt}% <replace>
  {}{}% <success><failure>
\makeatother

\graphicspath{{figures/}}

\def\MMSEPIC{\text{MMSE}}
\def\EP{\text{EP}}
\def\DEP{\text{DEP}}

\def\GEPNet{\text{GEP}}
\newcommand{\Times}[2]{${\text{#1}\times\text{#2}}$}
\newcommand{\SNR}[2]{${\text{SNR} #1 \text{#2}\;\text{dB}}$}

\newcommand{\SNRTRAIN}[2]{${\text{SNR}_{\text{train}} #1 \text{#2}\;\text{dB}}$}
\def\post{\textit{a posteriori }}
\def\prior{\textit{a priori }}
\def\GEP{GEPNet}
\def\TI{TI}

\begin{document}
\ifdefined \GramaCheck
  \newcommand{\CheckRmv}[1]{}
  \newcommand{\figref}[1]{Figure 1}%
  \newcommand{\tabref}[1]{Table 1}%
  \newcommand{\secref}[1]{Section 1}
  \newcommand{\algref}[1]{Algorithm 1}
  \renewcommand{\eqref}[1]{Equation 1}
\else
  \newcommand{\CheckRmv}[1]{#1}
  \newcommand{\figref}[1]{Fig.~\ref{#1}}%
  \newcommand{\tabref}[1]{Table~\ref{#1}}%
  \newcommand{\secref}[1]{Sec.~\ref{#1}}
  \newcommand{\algref}[1]{Algorithm~\ref{#1}}
  \renewcommand{\eqref}[1]{(\ref{#1})}
\fi
\newtheorem{theorem}{Theorem}
\newtheorem{proposition}{Proposition}
\newtheorem{assumption}{Assumption}
\newtheorem{definition}{Definition}
\newtheorem{condition}{Condition}
\newtheorem{property}{Property}
\newtheorem{remark}{Remark}
\newtheorem{lemma}{Lemma}
\newtheorem{corollary}{Corollary} 
%
% paper title
% Titles are generally capitalized except for words such as a, an, and, as,
% at, but, by, for, in, nor, of, on, or, the, to and up, which are usually
% not capitalized unless they are the first or last word of the title.
% Linebreaks \\ can be used within to get better formatting as desired.
% Do not put math or special symbols in the title.
\title{Graph Neural Network-Enhanced Expectation Propagation Algorithm for MIMO Turbo Receivers}
%
%
% author names and IEEE memberships
% note positions of commas and nonbreaking spaces ( ~ ) LaTeX will not break
% a structure at a ~ so this keeps an author's name from being broken across
% two lines.
% use \thanks{} to gain access to the first footnote area
% a separate \thanks must be used for each paragraph as LaTeX2e's \thanks
% was not built to handle multiple paragraphs
%

\author{Xingyu~Zhou,
        Jing~Zhang,
		Chao-Kai~Wen,~\IEEEmembership{Senior~Member,~IEEE,}
        Shi~Jin,~\IEEEmembership{Senior~Member,~IEEE,}
		and~Shuangfeng~Han,~\IEEEmembership{Senior~Member,~IEEE}% <-this % stops a space
\thanks{X.~Zhou, J.~Zhang, and S.~Jin are with the National Mobile
Communications Research Laboratory, Southeast University, Nanjing 210096, China
(e-mail: \protect \url{xy_zhou@seu.edu.cn}; jingzhang@seu.edu.cn; jinshi@seu.edu.cn).}% <-this % stops a space
\thanks{C.-K.~Wen is with Institute of Communications Engineering,
National Sun Yat-sen University, Kaohsiung 80424, Taiwan
(e-mail: chaokai.wen@mail.nsysu.edu.tw).}% <-this % stops a space
\thanks{S.~Han is with the China Mobile Research Institute, Beijing 100053, China (e-mail: \protect \url{hanshuangfeng@chinamobile.com}).}% <-this % stops a space
\thanks{This paper was presented in part at the 18th International Symposium on Wireless Communication Systems (ISWCS) \cite{zhouExtrinsicGraphNeural}.}}

% note the % following the last \IEEEmembership and also \thanks - 
% these prevent an unwanted space from occurring between the last author name
% and the end of the author line. i.e., if you had this:
% 
% \author{....lastname \thanks{...} \thanks{...} }
%                     ^------------^------------^----Do not want these spaces!
%
% a space would be appended to the last name and could cause every name on that
% line to be shifted left slightly. This is one of those "LaTeX things". For
% instance, "\textbf{A} \textbf{B}" will typeset as "A B" not "AB". To get
% "AB" then you have to do: "\textbf{A}\textbf{B}"
% \thanks is no different in this regard, so shield the last } of each \thanks
% that ends a line with a % and do not let a space in before the next \thanks.
% Spaces after \IEEEmembership other than the last one are OK (and needed) as
% you are supposed to have spaces between the names. For what it is worth,
% this is a minor point as most people would not even notice if the said evil
% space somehow managed to creep in.

% make the title area
\maketitle
% \vspace{-2.0cm}
% As a general rule, do not put math, special symbols or citations
% in the abstract or keywords.
\begin{abstract}
	% background: The role of AI in MIMO systems ok
	Deep neural networks (NNs) are considered a powerful tool for balancing the performance and complexity of multiple-input multiple-output (MIMO) receivers due to their accurate feature extraction, high parallelism, and excellent inference ability. 
	% background: GNN ok
	{Graph NNs (GNNs) have recently demonstrated outstanding capability} in learning enhanced message passing rules and have shown success in  overcoming the drawback of inaccurate Gaussian approximation of expectation propagation (EP)-based MIMO detectors.
	However, the application of the GNN-enhanced EP detector to MIMO turbo receivers is {underexplored and} non-trivial {due to the requirement of extrinsic information for iterative processing.}
	This paper proposes a GNN-enhanced EP algorithm for MIMO turbo receivers, which realizes the turbo principle of generating extrinsic information from the MIMO detector through a specially designed training procedure. 
	Additionally, an edge pruning strategy is designed to eliminate redundant connections in the original fully connected model of the GNN utilizing the correlation  information inherently from the EP algorithm. 
	Edge pruning reduces the computational cost dramatically and enables the network to focus more attention on the weights that are vital for performance. 
	Simulation results and complexity analysis indicate {that the proposed MIMO turbo receiver outperforms the EP turbo approaches} 
	by over 1 dB at the bit error rate of $10^{-5}$, 
	exhibits performance equivalent to  state-of-the-art receivers with 2.5 times shorter running time, and adapts to various scenarios. 
\end{abstract}

% \vspace{-0.5cm}
% Note that keywords are not normally used for peerreview papers.
\begin{IEEEkeywords}
	 Expectation propagation, graph neural network, MIMO turbo receiver, extrinsic information.
\end{IEEEkeywords}

% For peer review papers, you can put extra information on the cover
% page as needed:
% \ifCLASSOPTIONpeerreview
% \begin{center} \bfseries EDICS Category: 3-BBND \end{center}
% \fi
%
% For peerreview papers, this IEEEtran command inserts a page break and
% creates the second title. It will be ignored for other modes.
\IEEEpeerreviewmaketitle

\section{Introduction}
\IEEEPARstart{M}{ultiple}-input multiple-output (MIMO) has the potential to improve the link throughput by orders of magnitude and has become the key enabling technology 
for modern wireless communication systems that need to adapt to tremendous growth in transmission rate demand and network scale. 
To achieve the full benefits of MIMO technology, there is a strong demand for computationally efficient receiver designs, considering the increasing number of antennas used.

Several suboptimal linear detectors, such as the zero-forcing and linear minimum mean square error (LMMSE) algorithms, have been desirable among existing MIMO detectors \cite{yangFiftyYearsMIMO2015} because of their low computational cost.
However, compared with the maximum likelihood (ML) detector, substantial performance loss has restricted the application of linear detectors in future communication systems.
By contrast, the powerful sphere decoder (SD) \cite{studerSoftInputSoft2010} promises performance equivalent to the optimal ML. 
However, it is constrained to MIMO systems with a limited number of antennas 
because of the exponential worst-case complexity \cite{jaldenComplexitySphereDecoding2005}.

{Iterative detectors based on message passing (MP), specifically approximate MP (AMP) \cite{donohoMessagepassingAlgorithmsCompressed2009} and expectation propagation (EP){\cite{qiExpectationPropagationSignal2003,cespedesExpectationPropagationDetection2014,cespedesProbabilisticMIMOSymbol2018}}, have become promising strategies to approximate the ML detector with moderate complexity.} 
AMP is favored in addressing large-scale MIMO detection problems 
because its complexity is only quadratic to the system size.
However, {AMP is Bayes-optimal only when the channel matrix follows an independent and identically distributed (i.i.d.) sub-Gaussian distribution \cite{bayatiUniversalityPolytopePhase2015}, and it degrades significantly under realistic ill-conditioned channels.}
%% OAMP
{Orthogonal AMP (OAMP) \cite{maOrthogonalAMP2017} and vector AMP (VAMP) \cite{ranganVectorApproximateMessage2019} are powerful strategies that have been developed to address the limitations of AMP. 
They have been proven to achieve the Bayes-optimal performance for the general unitarily-invariant channel matrices under the large system limit \cite{takeuchiRigorousDynamicsExpectationPropagationBased2020}.
However, their performance tends to degrade in realistic finite-dimensional MIMO systems, %finite-sized
which are the main focus of this paper.\footnote{In this paper, our focus is on moderately sized  spatial-multiplexing MIMO systems, such as \Times{4}{4} and \Times{16}{16} configurations, rather than massive MIMO. 
This particular setup %has various potential benefits and 
is commonly found in current wireless standards and has received significant attention and research efforts  \cite{yangFiftyYearsMIMO2015}.}}

%% EP
EP relaxes the constraint on the channel matrix and outperforms the AMP detector over a wide range of MIMO channels by factorizing the posterior belief with Gaussian distributions \cite{minkaFamilyAlgorithmsApproximate}.
{Furthermore, EP demonstrates superior performance in small- or medium-sized MIMO systems as compared to OAMP/VAMP. 
This is attributed to EP's utilization of element-wise variance instead of the scalar variance employed in OAMP/VAMP \cite{takeuchiConvergenceOrthogonalVector2022}.} 
%%% turbo receiver
Iterative detection and decoding (IDD), or equivalently, MIMO turbo receiver, can be used to further improve detection accuracy,  which is a common practice in current communication systems. 
Moreover, EP-based turbo receivers have been widely applied{\cite{senstHowFrameworkExpectation2011, santosSelfTurboIterations2019,murillo-fuentesLowComplexityDoubleEPBased2021,sahinDoublyIterativeTurbo2019}}. 
%%% handcrafted 
However, EP detectors suffer a substantial gap with the ML performance because of the inaccuracy of Gaussian approximation in practical MIMO systems with high spatial correlation and strong interference. 

% DL-based MIMO detection -- more introduction
Recently, deep learning (DL) has demonstrated its remarkable capability of overcoming conventional challenges in wireless communications \cite{qinDeepLearningPhysical2019}. 
In particular, deep neural networks (NNs) offer the possibility to address existing gaps in iterative MIMO detectors  and promise great performance and efficiency in  receiver design \cite{heModelDrivenDeepLearning2020,samuelLearningDetect2019,khaniAdaptiveNeuralSignal2020,pratikREMIMORecurrentPermutation2021}.  
%% DetNet
The authors of \cite{samuelLearningDetect2019} developed a model-driven detection network (DetNet) by applying a projected gradient descent algorithm, and DetNet achieves the same accuracy as the AMP detector with enhanced robustness and lower complexity. 
However, DetNet requires substantial training data and performs poorly under high-order modulation or small-sized MIMO systems, thereby restricting its applications.  
%% OAMP-NET
Conventional MP-based detectors deteriorate severely in practical MIMO systems because of the failure of their prerequisites. 
An OAMP network (OAMPNet) was constructed in \cite{heModelDrivenDeepLearning2020} to compensate for the performance loss utilizing DL. 
However, the OAMPNet experiences performance degradation in real-world channels with strong correlations \cite{khaniAdaptiveNeuralSignal2020}.
More powerful NNs were introduced to enhance MP-based detectors by using highly parameterized models, including the MMNet in \cite{khaniAdaptiveNeuralSignal2020} and the recurrent equivariant MIMO detector in \cite{pratikREMIMORecurrentPermutation2021}. 
However, such schemes entail bulky detection networks with an excessive number of parameters to be trained.   
%% EP-NET
Furthermore, the EP detector was unfolded in \cite{sahinDoublyIterativeTurbo2019, zhangMetaLearningBasedMIMO2021} to derive a detection network with a few trainable damping factors. In this way, fast convergence can be achieved, and inefficient hand-crafted tuning processes can be avoided. 

%%  GNN 
Graph NNs (GNNs) provide an advanced technique for dealing with graph-structured data, and they have been widely applied to address the inference tasks of wireless communications \cite{heOverviewApplicationGraph2021}. 
% for MIMO detection
Recently, GNN-based MIMO detection \cite{scottiGraphNeuralNetworks2020, kosasih2022graph} has also attracted great attention because of the GNN's ability to enhance the MP solution by incorporating DL \cite{schmidLowComplexityNearOptimumSymbol2022}.  
The authors of \cite{scottiGraphNeuralNetworks2020} modeled the MIMO detection problem by the pair-wise Markov random field (MRF) and solved the corresponding maximum \post (MAP) inference task by learning an enhanced MP algorithm based on a GNN.
Furthermore, the authors of \cite{kosasih2022graph} developed a GNN-enhanced EP detector called {\GEP}, which introduced GNN to improve the posterior distribution approximation and exhibited a significant performance advantage over the EP and state-of-the-art NN-based detectors in uncoded systems.

However, 
the application of the {\GEP} to turbo iterative processing {is unexplored in \cite{kosasih2022graph},} which is not a trivial issue due to the requirement for extrinsic information,  
rather than \post probability (APP), to ensure a stable convergence. 
Furthermore, the dense connections in the fully connected (FC) MRF model result in a GNN with a deal of redundancy, which hinders the efficient implementation of the detector.

In this paper, we develop an \underline{ext}rinsic \underline{G}NN-aid \underline{EP} \underline{net}work (EXT-{\GEP}) for MIMO turbo receivers. 
We go beyond the design of GEPNet for uncoded systems \cite{kosasih2022graph} and follow the key idea of EP-based turbo receivers \cite{senstHowFrameworkExpectation2011, santosSelfTurboIterations2019,murillo-fuentesLowComplexityDoubleEPBased2021,sahinDoublyIterativeTurbo2019} to construct the turbo structure for GEPNet with soft inputs and soft outputs, enabling sufficient utilization of \prior information from the channel decoder to improve \post estimates.
We observe that the original {\GEP} \cite{kosasih2022graph} fails to generate reliable extrinsic information when applied to turbo iterative receiving, leading to poor performance. 
Hence, we customize a training scheme inspired by \cite{huangExtrinsicNeuralNetwork2021} to construct an %modified 
EXT-{\GEP} detector that satisfies the turbo principle \cite{tuchlerMinimumMeanSquared2002} of forwarding extrinsic information.
The training scheme addresses the limitations of the original {\GEP} and 
establishes a fine-tuned EXT-{\GEP} that can be integrated into the developed turbo structure to realize IDD with great flexibility and  remarkable performance.
{We also reduce the computational complexity of the original {\GEP} by designing an edge pruning scheme to simplify the GNN operations to a large extent while still maintaining excellent performance.}

The contributions of this paper are summarized as follows:
\begin{itemize}
	\item \textit{Design of an EXT-{\GEP}-based turbo receiver.} 
	 %ok
	Through intensive simulation studies, we discovered that the original GEPNet cannot generate desired extrinsic outputs via the conventional strategy used in the MAP detector. 
	To address this issue, we design an open-loop training scheme to derive an EXT-{\GEP} that produces the extrinsic outputs required by the turbo procedure.
	This scheme is formulated based on a preliminary NN and the requirement for extrinsic information \cite{tuchlerMinimumMeanSquared2002}, i.e., not coupling with the priors, 
	to obtain target extrinsic log-likelihood ratio (LLR) samples as labels for training the final EXT-{\GEP}. 
	The fine-tuned network can be plugged into the developed turbo structure to construct the EXT-GEPNet-based turbo receiver, which effectively overcomes the correlation problems when using the original GEPNet by preventing the same information from being counted twice. 
	The proposed scheme is more flexible than directly training through the IDD procedure, as it does not rely on the choice of channel codes. 
	\item \textit{Edge pruning to reduce the complexity.} 
	To relieve the complexity of the MP process, we remove redundant edges in the latent FC graph of the GNN.
	{Unlike existing NN pruning schemes that simply drop the network's weights with small magnitudes \cite{buchbergerPruningQuantizingNeural2021,schmidLowComplexityNearOptimumSymbol2022},} we perform edge pruning utilizing the correlation information among the to-be-estimated variables inherent in the EP iterations, {inspired by \cite{scotti2020graph}}. 
	The proposed network can complete the detection with lower computational cost and even achieve performance gain after pruning due to the reduction of ineffective connections and avoidance of overfitting. 
	
    \item \textit{Comprehensive performance and complexity evaluation}. We validate our scheme with a wide range of numerical simulations under different scenarios that
	benchmark against a series of baselines. 
	We also analyze the computational complexity of the proposed turbo receiver. 
	Simulation results show that the proposed receiver has significant gain over {the turbo approaches supported by EP and the original {\GEP}} and achieves comparable or even better performance with substantially lower running time than the single tree-search SD (STS-SD)-based receiver \cite{studerSoftInputSoft2010}.  
	{Furthermore, the proposed receiver adapts well to different channels and channel codes {and exhibits robustness to channel estimation errors}.}
	
\end{itemize}

\textit{Notations:} 
Boldface letters denote column vectors or matrices. 
$\mathbf{A}^T$ and $\mathbf{A}^{\dagger} = (\mathbf{A}^T\mathbf{A})^{-1}\mathbf{A}^T$ represent the transpose and pseudo-inverse of matrix $\mathbf{A}$, respectively. 
${\mathbf{I}}_N$ is an identity matrix of size $N$, and
$\mathbf{0}$ is a zero matrix. 
$\delta(\cdot)$, ${\mathbb{E}}[\cdot]$, and $\left \| \cdot \right \|$ denote the Dirac delta function, expectation operation, and Euclidean norm, respectively. 
The set $[K]={1,2,\dots,K}$ contains all nonnegative integers up to $K$.
Finally, ${\cal N}( {z;\mu ,{\sigma ^2}} )$ denotes real-valued Gaussian random variables with mean $\mu $ and variance $\sigma^2$.

%%%%%%%%%%%%%%%%%%%%%%%%%%%%%%%%%%%%%%%%%%%%%%%%%%%%%%%%%%
\section{System Model and Algorithm Review} \label{sec:sys_model}
In this section, the system model of the IDD problem is formulated first. 
Then, the EP-based turbo receiver is reviewed to obtain a clear understanding of the proposed scheme.

\subsection{System Model}
The considered MIMO system on the basis of bit-interleaved coded modulation (BICM) consists of $N_{\rm t}$ transmit (Tx) antennas and $N_{\rm r}$ receive (Rx) antennas, with $N_{\rm r} \geq N_{\rm t}$.  
\figref{fig:sys_model} depicts the system with a block diagram, which includes 
a MIMO transmitter and a MIMO turbo receiver. % MIMO turbo receiver
At the transmitter, the channel encoder first converts the binary word, $\mathbf{b} \in \{0,1\}^{N_{\rm b}}$ with $N_{\rm b}$ as the number of message bits in a word, into the coded bits with code rate $R_{\rm c}=N_{\rm b}/N_{\rm c}$. 
Then, interleaving is performed to the coded bits, thereby yielding the codeword $\mathbf{c}$ of length $N_{\rm c}$.
The codeword $\mathbf{c}$ is then partitioned into $N_{\rm s}$ subvectors of length $N_{\rm t}\tilde{Q}$ and modulated into symbol vectors with a complex quadrature amplitude modulation (QAM) constellation $\tilde{\mathcal{A}}$ of size $|\tilde{\mathcal{A}}|=\tilde{M}$, where $\tilde{Q}$ is the number of bits per complex symbol and $\tilde{Q}=\log_2\tilde{M}$. 
The symbol vectors are transmitted over the wireless channel, which is assumed to be unchanged in a time slot, and the received real-valued signal can be represented as
\CheckRmv{
\begin{equation}
	\mathbf{y} =\mathbf{Hx}+\mathbf{w},
	\label{eq:sys_model_real}
\end{equation}
} 
where $\mathbf{x} \in \mathcal{A}^{K}$ with $K=2N_{\rm t}$ is the equivalent real-valued transmitted vector in a time slot. The real-valued constellation $\mathcal{A}$ has a cardinality of $|\mathcal{A}|\triangleq M=\sqrt{\tilde{M}}$ and average energy of $E_{\rm s}$.  
The channel matrix $\mathbf{H}\in \mathbb{R}^{N\times K}$, with $N=2N_{\rm r}$ and its columns $\mathbf{h}_k, k\in [K]$ normalized to unit energy, is supposed to be known at the receiver without special illustrations.  %  unchanged in a time slot
${\mathbf{w}}$ is the additive white Gaussian noise vector with zero mean and element-wise noise variance $\sigma_w^2$.

\CheckRmv{
\begin{figure}[t]
	\centering
	\includegraphics[width=3.5in]{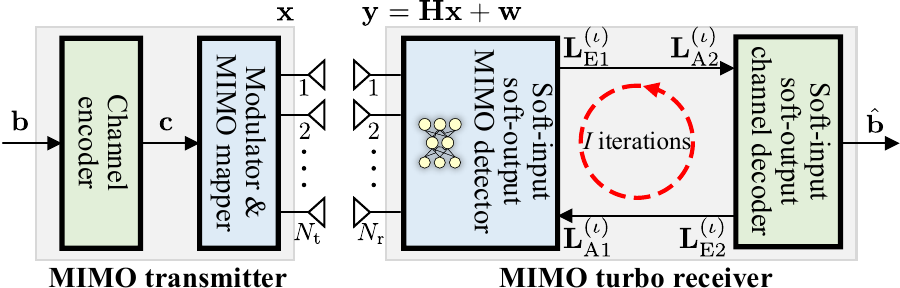}
	\caption{{Block diagram of a MIMO system based on BICM \cite{studerSoftInputSoft2010}. 
	The MIMO turbo receiver iteratively exchanges soft information between the MIMO detector, which combines model-based algorithms with NNs, and the channel decoder.}
	}
	\label{fig:sys_model}
\end{figure}
}

The posterior probability density function (PDF) of the transmitted symbol vector $\mathbf{x}$ given the observations $\mathbf{y}$ yields 
\CheckRmv{
\begin{equation}
	p(\mathbf{x} | \mathbf{y})=\frac{p(\mathbf{y} | \mathbf{x})p(\mathbf{x}) }{p(\mathbf{y})} \propto \underbrace{\mathcal{N}\left(\mathbf{y}; \mathbf{H} \mathbf{x}, \sigma_w^{2} \mathbf{I}_{N}\right)}_{p(\mathbf{y} | \mathbf{x})} \underbrace{\prod_{k=1}^{K} {p_{\rm A1}}(x_{k})}_{p(\mathbf{x})},
	\label{eq:post}
\end{equation}
}
where ${p_{\rm A1}}(x_{k})$ is the \prior PDF of ${x}_k$. 
In the first turbo iteration (TI), the prior is initialized as ${p_{\rm A1}^{(1)}}(x_k)=\frac{1}{M}\sum_{x\in \mathcal{A}}\delta(x_k-x)$ assuming equiprobable transmitted symbols. 
The direct calculation of $p(\mathbf{x}|\mathbf{y})$ 
involves a high-dimensional integral, which is generally intractable. Therefore, Bayesian inference techniques (e.g., EP) are commonly used to compute an approximation $q(\mathbf{x})$ 
for the optimal solution.
The IDD technique, which is referred to as the turbo receiver in this paper, can  improve the accuracy of the approximation further. 
As illustrated in \figref{fig:sys_model}, the soft-input soft-output signal detector and channel decoder in the turbo structure iteratively exchange reliability information on the same set of coded bits in the form of LLRs.
These LLRs correspond to extrinsic probabilities
to ensure the convergence and stability of the receiver.
Moreover, the improved prior information can be utilized when the turbo procedure begins, that is, ${p_{\rm A1}^{(\iota)}}(x_{k})$ can be constructed on the basis of the feedback from the channel decoder instead of the equiprobable assumption, where $\iota$ is the index of the TI. 
The iterative process proceeds for a maximum number of $I$ iterations and finally outputs the estimated message bits $\hat{\mathbf{b}}$. 

% ok
The extrinsic LLR for $c_{k,i}$, which is the $i$-th bit mapped to symbol $x_k$, can be computed on the basis of the extrinsic PDF ${p_{\rm E1}^{(\iota)}}(x_k|\mathbf{y})$ estimated by the detector in each TI as
\CheckRmv{
\begin{equation}
	{L_{\mathrm{E1}}^{(\iota)}(c_{k,i})} \triangleq \log \frac{\sum_{x_k \in \mathcal{A}_{k,i}^{(1)}} {p_{\rm E1}^{(\iota)}}(x_{k} | \mathbf{y})}{\sum_{x_k \in \mathcal{A}_{k,i}^{(0)}} {p_{\rm E1}^{(\iota)}}(x_{k} | \mathbf{y})},
  \label{eq:ext_llr}
\end{equation}
}
where $\mathcal{A}_{k,i}^{(1)}$ and $\mathcal{A}_{k,i}^{(0)}$ denote the subsets of constellation $\mathcal{A}$, in which $c_{k,i}$ is equal to 1 and 0, respectively.
% ok
The vector ${\mathbf{L}_{\rm E1}^{(\iota)}}$, which contains all extrinsic LLR values from the detector, is further de-interleaved to derive ${\mathbf{L}_{\rm A2}^{(\iota)}}$, which is delivered to the channel decoder as the \prior LLRs.
The decoder computes the extrinsic LLRs on the coded bits, indicated as ${\mathbf{L}_{\rm E2}^{(\iota)}}$, utilizing the \prior LLRs ${\mathbf{L}_{\rm A2}^{(\iota)}}$.    
{For the $(\iota+1)$-th TI}, the extrinsic LLRs generated by the decoder are interleaved to derive ${\mathbf{L}_{\rm A1}^{(\iota+1)}}$, sent back to the detector,  and mapped again to the updated \prior PDF: 
\CheckRmv{
	\begin{equation}
		{p_{\rm A1}^{(\iota+1)}}(x_{k})= \prod_{i=1}^{Q} \frac{\exp{(c_{k,i}{L_{\mathrm{A1}}^{(\iota+1)}(c_{k,i})})}}{1+\exp{({L_{\mathrm{A1}}^{(\iota+1)}(c_{k,i})})}},
    \label{eq:prior}
	\end{equation}
} 
where $Q=\tilde{Q}/2$ denotes the number of bits in a real-valued symbol. 

\subsection{EP-based Turbo Receiver} \label{sec:ep}
EP is a Bayesian inference method that approximates the desired distribution by a function within the exponential families \cite{seegerExpectationPropagationExponential2009}.{\footnote{{For a comprehensive review of the exponential families and their properties, we recommend the readers to refer to \cite{wainwrightGraphicalModelsExponential}.}}}
Specifically, the EP-based MIMO detector computes a Gaussian approximation $q(\mathbf{x})$
for the posterior belief $p(\mathbf{x} | \mathbf{y})$ in \eqref{eq:post}, which is achieved by replacing the non-Gaussian factors (discrete priors) in \eqref{eq:post} with unnormalized Gaussians iteratively:
\CheckRmv{
	 \begin{align} 
		q^{(\iota,t)}(\mathbf{x})  \propto &\mathcal{N}(\mathbf{y}; \mathbf{H} \mathbf{x}, \sigma_w^{2} \mathbf{I}_{N}) \cdot \prod_{k=1}^{K} \exp{\big(\gamma_{k}^{(\iota,t-1)} x_{k}-\frac{1}{2} \lambda_{k}^{(\iota,t-1)} x_{k}^{2}\big)} \nonumber \\
		 \propto & \mathcal{N}\big(\mathbf{x}; \mathbf{H}^{\dagger} \mathbf{y}, \sigma_w^{2}(\mathbf{H}^{T} \mathbf{H})^{-1}\big)\nonumber \\
		&\cdot \mathcal{N}\big(\mathbf{x};(\boldsymbol{\lambda}^{(\iota,t-1)})^{-1} \boldsymbol{\gamma}^{(\iota,t-1)},(\boldsymbol{\lambda}^{(\iota,t-1)})^{-1}\big) \nonumber\\
		 \propto & \mathcal{N}\big(\mathbf{x}; \boldsymbol{\mu}^{(\iota,t)}, \boldsymbol{\Sigma}^{(\iota,t)}\big), 
	\end{align}
}
where the superscript {${(\iota,t)}$ denotes the $t$-th EP iteration within the $\iota$-th TI}. 
{$\gamma_{k}^{(\iota, t)} \in \mathbb{R}$ and $\lambda_{k}^{(\iota, t)} \in \mathbb{R}^{+}$ denote the natural parameters of the 
approximating function, 
constituting the natural mean vector $\boldsymbol{\gamma}^{(\iota, t)}=[\gamma_{1}^{(\iota, t)},\dots,\gamma_{K}^{(\iota, t)}]^{T}$ and precision matrix $\boldsymbol{\lambda}^{(\iota, t)}={\rm diag}([\lambda_{1}^{(\iota, t)},\dots,\lambda_{K}^{(\iota, t)}])$ \cite{seegerExpectationPropagationExponential2009}}.
{In the first TI}, parameters $\gamma_{k}^{(\iota, t)}$ and $\lambda_{k}^{(\iota, t)}$ are initialized as $\gamma_{k}^{(1,0)}=0$ and $\lambda_{k}^{(1,0)}=1/E_{\rm s}, k\in [K]$, respectively, 
and then iteratively updated following the moment matching condition as $\mathbb{E}_{q(\mathbf{x})}=\mathbb{E}_{p(\mathbf{x}|\mathbf{y})}$ \cite{cespedesExpectationPropagationDetection2014}. The mean $\boldsymbol{\mu}^{(\iota, t)}$ and covariance $\boldsymbol{\Sigma}^{(\iota, t)}$ of $q^{(\iota, t)}(\mathbf{x})$ are computed using the Gaussian product lemma \cite{rasmussenGaussianProcessesMachine2004} as follows:
\CheckRmv{
  \begin{subequations}
    \begin{align}
    	\boldsymbol{\Sigma}^{(\iota, t)}&=\big(\sigma_w^{-2}{\mathbf{H}}^{T} {\mathbf{H}}+\boldsymbol{\lambda}^{(\iota, t-1)}\big)^{-1}, \label{eq:sigma} \\
		\boldsymbol{\mu}^{(\iota, t)}&=\boldsymbol{\Sigma}^{(\iota, t)}\big(\sigma_w^{-2}{\mathbf{H}}^{T} \mathbf{y}+\boldsymbol{\gamma}^{(\iota, t-1)}\big).
	\end{align}
	\label{eq:lmmse}%
  \end{subequations} 
} 
Moreover, EP calculates the marginal distribution of $q^{(\iota, t)}(\mathbf{x})$ by viewing the covariance $\boldsymbol{\Sigma}^{(\iota, t)}$ as a diagonal matrix to reduce complexity, that is, $q^{(\iota, t)}(x_k) \propto \mathcal{N}(x_k;\mu_k^{(\iota, t)},\Sigma_k^{(\iota, t)}), k\in [K]$, where $\mu_k^{(\iota, t)}$ is the $k$-th element of $\boldsymbol{\mu}^{(\iota, t)}$, and $\Sigma_k^{(\iota, t)}$ is the $k$-th element of the main diagonal in $\boldsymbol{\Sigma}^{(\iota, t)}$. This strategy, which uses the product of independent Gaussian functions to approximate $p(\mathbf{x} | \mathbf{y})$ \cite{kosasih2022graph}, 
ignores the off-diagonal elements in the covariance $\boldsymbol{\Sigma}^{(\iota, t)}$ and results in information loss. 

\CheckRmv{
	\begin{figure}[t]
		\centering
		\setlength{\abovecaptionskip}{-0.1cm}
		\setlength{\belowcaptionskip}{-0.0cm}
		\includegraphics[width=3.5in]{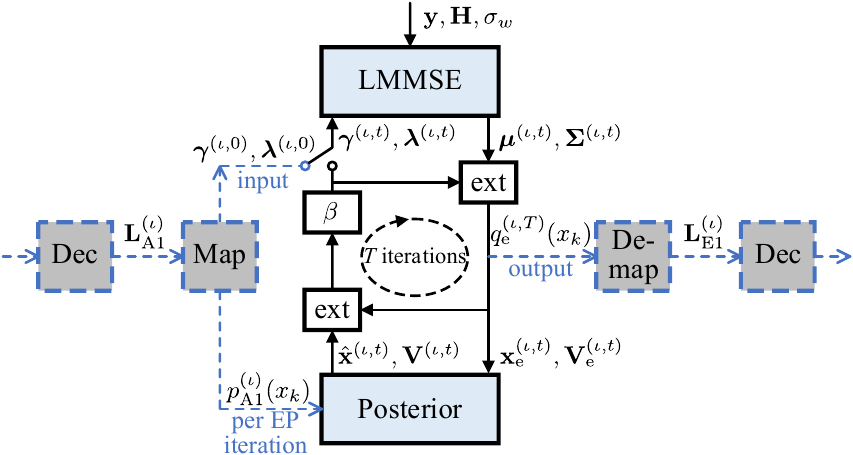}
		\caption{{Visual representation of the EP-based turbo receiver at the $\iota$-th turbo iteration. The black solid lines indicate the EP iterations, while the blue dash lines indicate the interaction with the channel decoder (Dec).}} 
		\label{fig:ep}
	\end{figure}
}

% above is ok
\figref{fig:ep} presents the block diagram of the {EP detector at the $\iota$-th TI}, which contains a LMMSE module and a nonlinear Posterior module and bears a turbo structure. The LMMSE module is dedicated to the computation of $\boldsymbol{\mu}^{(\iota, t)}$ and $\boldsymbol{\Sigma}^{(\iota, t)}$ in \eqref{eq:lmmse}. 
Then, the extrinsic marginal distribution is derived by the ``ext'' operation after LMMSE to decorrelate the output and the input as follows: 
\CheckRmv{
	\begin{align}
		q_{\rm e}^{(\iota, t)}\left(x_{k}\right) &= \frac{q^{(\iota, t)}\left(x_{k}\right)}{\exp{\big(\gamma_{k}^{(\iota, t-1)} x_{k}-\frac{1}{2} \lambda_{k}^{(\iota, t-1)} x_{k}^{2}\big)}} \nonumber\\
		&\propto \mathcal{N}\big(x_{k}; x_{{\rm e}, k}^{(\iota, t)}, v_{{\rm e}, k}^{(\iota, t)}\big), \quad k\in [K], 
		\label{eq:ext_prop}
	\end{align}
}
where $x_{{\rm e}, k}^{(\iota, t)}$ and $v_{{\rm e}, k}^{(\iota, t)}$ are the $k$-th element of the mean vector $\mathbf{x}_{\rm e}^{(\iota, t)}$ and the main diagonal in the covariance matrix $\mathbf{V}_{\rm e}^{(\iota, t)}$, respectively, and we yield
\CheckRmv{
	\begin{subequations}
		\begin{align}
			v_{{\rm e}, k}^{(\iota, t)} &=\frac{\Sigma_{k}^{(\iota, t)}}{1-\Sigma_{k}^{(\iota, t)} \lambda_{k}^{(\iota, t-1)}}, \\ 
			x_{{\rm e}, k}^{(\iota, t)} &=v_{{\rm e}, k}^{(\iota, t)}\left(\frac{\mu_{k}^{(\iota, t)}}{\Sigma_{k}^{(\iota, t)}}-\gamma_{k}^{(\iota, t-1)}\right).
		\end{align}%
		\label{eq:cavity}%
	\end{subequations}
}
Subsequently, the mean vector $\mathbf{x}_{\rm e}^{(\iota, t)}$ and diagonal covariance matrix $\mathbf{V}_{\rm e}^{(\iota, t)}$ of the extrinsic distribution are delivered to the Posterior module for further processing.

{The Posterior module combines the  
\prior PDF $p_{\rm A1}^{(\iota)}(x_k)$ with the extrinsic distribution to derive the estimated \post distribution as \cite{senstHowFrameworkExpectation2011, santosSelfTurboIterations2019,murillo-fuentesLowComplexityDoubleEPBased2021,sahinDoublyIterativeTurbo2019}:
\CheckRmv{
	\begin{equation}
		\hat{p}^{(\iota, t)}(x_k
		) \propto q_{\rm e}^{(\iota, t)}(x_{k}){p_{\rm A1}^{(\iota)}}(x_k),
		\label{eq:est_post}
	\end{equation}
}
where the \prior PDF $p_{\rm A1}^{(\iota)}(x_k)$ is uniform for $\iota=1$ and non-uniform for $\iota\geq 2$ mapped by the \prior LLRs ${\mathbf{L}_{\rm A1}^{(\iota)}}$ from the channel decoder according to \eqref{eq:prior}. 
The \post mean $\hat{x}_k^{(\iota, t)}$ and variance $v_k^{(\iota, t)}$ are computed as:}
\CheckRmv{
  \begin{subequations}
		\begin{align}
			\hat{x}_{k}^{(\iota, t)} &=\sum_{a_m \in \mathcal{A}}a_m  {\hat{p}^{(\iota, t)}(x_k=a_m)}, \\ 
			v_{k}^{(\iota, t)} &=\sum_{a_m \in \mathcal{A}}\big(x_{k}-\hat{x}_{k}^{(\iota, t)}\big)^{2}{\hat{p}^{(\iota, t)}(x_k=a_m)}.
		\end{align}%
		\label{eq:ep_post}%
	\end{subequations}
}
$T$ is denoted as the number of EP iterations, and the pair $(\boldsymbol{\gamma}^{(\iota, t)},\boldsymbol{\lambda}^{(\iota, t)})$ is updated when $t<T$ so that
\CheckRmv{
	\begin{align}
		&\prod_{k=1}^{K}\exp{\big(\gamma_{k}^{(\iota, t)} x_{k}-\frac{1}{2} \lambda_{k}^{(\iota, t)} x_{k}^{2}\big)} \nonumber\\
		\propto& \mathcal{N}\big(\mathbf{x};(\boldsymbol{\lambda}^{(\iota, t)})^{-1} \boldsymbol{\gamma}^{(\iota, t)},(\boldsymbol{\lambda}^{(\iota, t)})^{-1}\big)  
		\propto \frac{\mathcal{N}\big(\mathbf{x}; \hat{\mathbf{x}}^{(\iota, t)}, \mathbf{V}^{(\iota, t)}\big)}{\mathcal{N}\big(\mathbf{x}; \mathbf{x}_{\rm e}^{(\iota, t)}, \mathbf{V}_{\rm e}^{(\iota, t)}\big)},
		\label{eq:gauss_div}
	\end{align}
}
where $\hat{\mathbf{x}}^{(\iota, t)}=[\hat{x}_{1}^{(\iota, t)},\dots,\hat{x}_{K}^{(\iota, t)}]^{T}$ and $\mathbf{V}^{(\iota, t)}={\rm diag}([v_{1}^{(\iota, t)},\dots,v_{K}^{(\iota, t)}])$. 
A solution to \eqref{eq:gauss_div} is: %given by 
\CheckRmv{
	\begin{subequations}
		\begin{align}
			&\boldsymbol{\lambda}^{(\iota, t)}=\left(\mathbf{V}^{(\iota, t)}\right)^{-1}-\left(\mathbf{V}_{\rm e}^{(\iota, t)}\right)^{-1}, \label{eq:lambda} \\
			&\boldsymbol{\gamma}^{(\iota, t)}=\left(\mathbf{V}^{(\iota, t)}\right)^{-1} \hat{\mathbf{x}}^{(\iota, t)}-\left(\mathbf{V}_{\rm e}^{(\iota, t)}\right)^{-1} \mathbf{x}_{\rm e}^{(\iota, t)},
		\end{align}%
		\label{eq:post_ext}%
	\end{subequations}
}
which is implemented by the ``ext'' operation after the Posterior module.
Notably, the update in \eqref{eq:lambda} can result in a negative ${\lambda}_k^{(\iota, t)}$, which is unreasonable {and should be discarded} because $\boldsymbol{\lambda}^{(\iota, t)}$ is an inverse variance term. 
{Therefore, we adopt the approach from \cite{cespedesExpectationPropagationDetection2014} to ensure numerical stability, where we retain} ${\lambda}_k^{(\iota, t)}={\lambda}_k^{(\iota, t-1)}$ and ${\gamma}_k^{(\iota, t)}={\gamma}_k^{(\iota, t-1)}$ when ${\lambda}_k^{(\iota, t)}<0$. 
{Additionally, we apply a damping technique \cite{cespedesExpectationPropagationDetection2014,seegerExpectationPropagationExponential2009} to smooth the update by using a convex combination of the previous value:}
\CheckRmv{
	\begin{subequations}
	\begin{align}
		\boldsymbol{\lambda}^{(\iota, t)}&\leftarrow\beta\boldsymbol{\lambda}^{(\iota, t)}+(1-\beta)\boldsymbol{\lambda}^{(\iota, t-1)},\\
		\boldsymbol{\gamma}^{(\iota, t)}&\leftarrow\beta\boldsymbol{\gamma}^{(\iota, t)}+(1-\beta)\boldsymbol{\gamma}^{(\iota, t-1)},
	\end{align}%
	\label{eq:damping}%
	\end{subequations}
}
where $\beta \in [0,1]$ is a damping factor. 
{This approach ensures a robust algorithm with improved stability and convergence properties.}
The updated pair $(\boldsymbol{\gamma}^{(\iota, t)},\boldsymbol{\lambda}^{(\iota, t)})$ is delivered to the LMMSE module for the next EP iteration. 

The EP algorithm finishes when the maximum number of iterations $T$ is reached.  {Extrinsic LLRs {$\mathbf{L}_{\rm E1}^{(\iota)}$} are {demapped} from the extrinsic distribution {$q^{(\iota,T)}_{\rm e}(x_k)$} via \eqref{eq:ext_llr} at the final iteration and delivered to the channel decoder, which outputs the estimated bits $\hat{\mathbf{b}}$ when $\iota=I$ or new \prior LLRs $\mathbf{L}_{\rm A1}^{(\iota+1)}$ when $\iota<I$ for subsequent TIs.
These \prior LLRs are mapped to the updated \prior PDF $p_{\rm A1}^{(\iota+1)}(x_k)$. 
The detector then computes the mean and variance of $p_{\rm A1}^{(\iota+1)}(x_k)$ as: 
\CheckRmv{
	\begin{subequations}
		\begin{align}
			\hat{x}_{{\rm A1},k}^{(\iota+1)} &=\sum_{a_m \in \mathcal{A}}a_m  p_{\rm A1}^{(\iota+1)}(x_k=a_m), \\ 
			v_{{\rm A1},k}^{(\iota+1)} &=\sum_{a_m \in \mathcal{A}}\big(x_{k}-\hat{x}_{{\rm A1},k}^{(\iota+1)}\big)^{2}p_{\rm A1}^{(\iota+1)}(x_k=a_m),
		\end{align}%
		\label{eq:prior_mean_var}%
	\end{subequations}
}
and updates the initial pair $(\boldsymbol{\gamma}^{(\iota+1,0)},\boldsymbol{\lambda}^{(\iota+1,0)})$ for EP at the ${(\iota+1)}$-th %($\iota=2,\dots,I$) 
{\TI} as:} 
\CheckRmv{
	\begin{equation}
		\boldsymbol{\lambda}^{(\iota+1,0)} \leftarrow ({\mathbf{V}^{(\iota+1)}_{\mathrm{A1}}})^{-1},\quad  \boldsymbol{\gamma}^{(\iota+1,0)}\leftarrow\boldsymbol{\lambda}^{(\iota+1,0)}{\hat{\mathbf{x}}^{(\iota+1)}_{\mathrm{A1}}},
		\label{eq:init_pair}
	\end{equation}
}
{with $\hat{\mathbf{x}}^{(\iota+1)}_{\mathrm{A1}}=[\hat{x}_{{\rm A1},1}^{(\iota+1)},\dots,\hat{x}_{{\rm A1},K}^{(\iota+1)}]^{T}$ and $\mathbf{V}^{(\iota+1)}_{\mathrm{A1}}={\rm diag}([v_{{\rm A1},1}^{(\iota+1)},\dots,v_{{\rm A1},K}^{(\iota+1)}])$.}

%%%%%%%%%%%%%%%%%%%%%%%%%%%%%%%%%%%%%%%%%%%%%%%%%%%%%%%%%%
\CheckRmv{
	\begin{figure*}[t]
	\setlength{\abovecaptionskip}{-0.1cm}
	\setlength{\belowcaptionskip}{-0.0cm}
		\centering
		\includegraphics[width=4.5in]{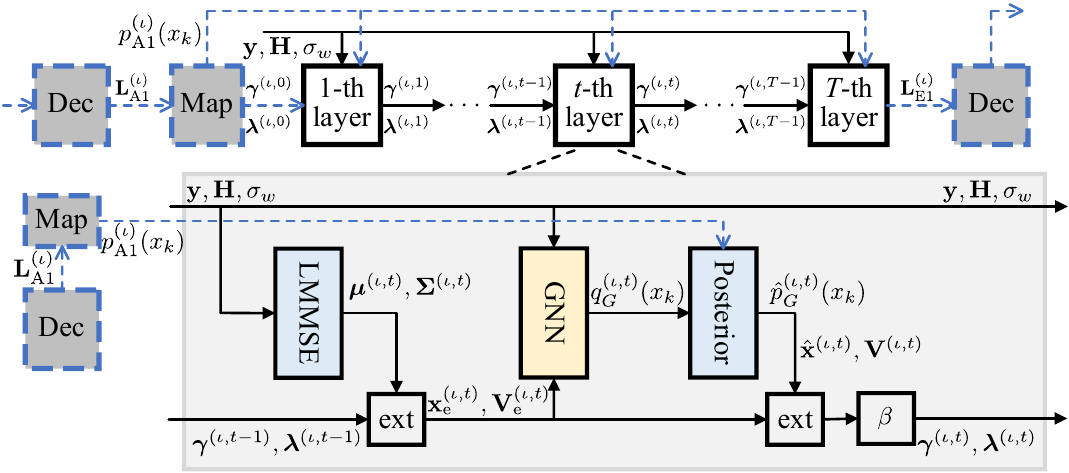}
		\caption{{Visual representation of the proposed turbo receiver structure for {\GEP} at the $\iota$-th turbo iteration. The black solid lines depict the unfolding structure of {\GEP}, while the blue dash lines depict the interaction with the channel decoder (Dec).}} %  
		\label{fig:gnn-ep}
	\end{figure*}
}
\section{EXT-{\GEP} for Turbo Receiver} \label{sec:GNN-EP for Turbo}
This section presents the details of the proposed EXT-GEPNet. 
We first introduce the {\GEP} detector \cite{kosasih2022graph} to improve the posterior distribution approximation of EP.
{Then, we present the designed turbo structure for {\GEP} and the customized training scheme used to derive EXT-{\GEP}.} 
Finally, we introduce the edge pruning method to simplify the GNN operations.

%%%%%%%%%%%%%%%%%%%%%%%%%%%%%%%%%%%%%%%%%%%%%%%%%%%%%%%%%%
\subsection{GNN-Enhanced EP Detector} \label{sec:GNN-EP}

% ok
EP factorizes the target APPs with Gaussian distributions to avoid intractable calculations.
% ok
However, the Gaussian approximation of the posterior belief becomes inaccurate in practical scenarios. 
For example, in a MIMO system with the number of Tx antennas close to that of the Rx antennas, where strong interference is extremely detrimental, the residual noise that can be viewed as Gaussian sharply decreases. 
This situation leads to a significant performance gap between the EP detector and the ML detector. 
To solve this limitation of EP, the GEPNet detector \cite{kosasih2022graph} was proposed, as shown in \figref{fig:gnn-ep} with black solid lines. 
% ok
{The {\GEP} structure is obtained by first unfolding the EP iterations into layers and then adding a GNN module to each layer while retaining the LMMSE and Posterior modules.}\footnote{We denote an EP iteration as a layer in the rest of this paper for consistency.}
The GNN module is inserted between the two original EP's modules and is used to provide an improved estimated distribution of the transmitted symbols over the constellation $\mathcal{A}$ based on the prior knowledge $q_{\rm e}^{(t)}(x_k)$ from \eqref{eq:ext_prop}.\footnote{{The GEPNet employs the same model parameters for different turbo iterations in the proposed turbo receiver. Therefore, in this section,} we omit the superscript $\iota$ of the turbo iteration index  for simplicity.}
% ok 
Subsequently, we elaborate the GNN module \cite{scottiGraphNeuralNetworks2020}.

The GNN module (\figref{fig:gnn}) provides a framework used for capturing the structured dependency of the transmitted variables $\mathbf{x}$ by combining DL into the MP on the pair-wise MRF model
\cite{scottiGraphNeuralNetworks2020}, where each variable is denoted as a node, and each pair of two nodes is linked by an edge. 
The nodes and edges in the MRF are represented by circles and squares in \figref{fig:gnn}, respectively. 
The prior knowledge $q_{\rm e}^{(t)}(x_k)$, 
%from the LMMSE module of EP, 
which is characterized by the mean $x_{{\rm e},k}^{(t)}$ and variance $v_{{\rm e},k}^{(t)}$, is incorporated into the node attribute as $\mathbf{a}_k^{(t)}=[x_{{\rm e},k}^{(t)},v_{{\rm e},k}^{(t)}]$.

Furthermore, the nodes and edges of the MRF are both characterized by their feature vectors, which are denoted as $\mathbf{u}_k^{(l)}$ and $\mathbf{f}_{jk}$, respectively, where $l$ is the iteration index of the MP and $j\in [K] \backslash \{k\}$. 
The node feature vector $\mathbf{u}_k^{(l)}$ of size $N_{\rm u}$ encodes the probabilistic information about the variable $x_k$ that corresponds to the self potential of the MRF, 
{given by \cite[Eq. (5)]{scottiGraphNeuralNetworks2020} as}
\CheckRmv{
	\begin{equation*}
		{\phi\left(x_{k}\right)=\exp \left(\frac{1}{\sigma_w^{2}} \mathbf{y}^{T} \mathbf{h}_{k} x_{k}-\frac{1}{2\sigma_w^{2}} \mathbf{h}_{k}^{T} \mathbf{h}_{k} x_{k}^{2}\right) p_{\rm A1}\left(x_{k}\right).}
	\end{equation*}
}
In particular, $\mathbf{u}_k^{(l)}$ is initialized as $\mathbf{u}_{k}^{(0)}=\mathbf{W}_{1} \cdot\left[\mathbf{y}^{T} \mathbf{h}_{k}, \mathbf{h}_{k}^{T} \mathbf{h}_{k}, \sigma_w^{2}\right]^{T}+\mathbf{b}_{1}$ and iteratively updated in the MP process, where $\mathbf{W}_1 \in \mathbb{R}^{N_{\rm u} \times 3}$ is a learnable weight matrix, and $\mathbf{b}_1\in \mathbb{R}^{N_{\rm u}}$ is a learnable bias vector. 
The edge feature vector $\mathbf{f}_{jk}=[\mathbf{h}_k^{T}\mathbf{h}_j,\sigma_w^2]$ contains the pair potential between nodes $x_j$ and $x_k$,  
{given by \cite[Eq. (6)]{scottiGraphNeuralNetworks2020} as}
\CheckRmv{
	\begin{equation*}
		{\psi\left(x_{k}, x_{j}\right)=\exp \left(-\frac{1}{\sigma_w^{2}} \mathbf{h}_{k}^{T} \mathbf{h}_{j} x_{k} x_{j}\right).}
	\end{equation*}
}

Three steps are involved in the MP process of the GNN: propagation, aggregation, and readout. 
The first two steps are implemented at each MP iteration $l$, whereas the readout  is conducted at the final iteration $L$ to make the inference. 
\CheckRmv{
	\begin{figure}[tbp]
	\setlength{\abovecaptionskip}{-0.1cm}
	\setlength{\belowcaptionskip}{-0.0cm}
		\centering
		\includegraphics[width=2.8in]{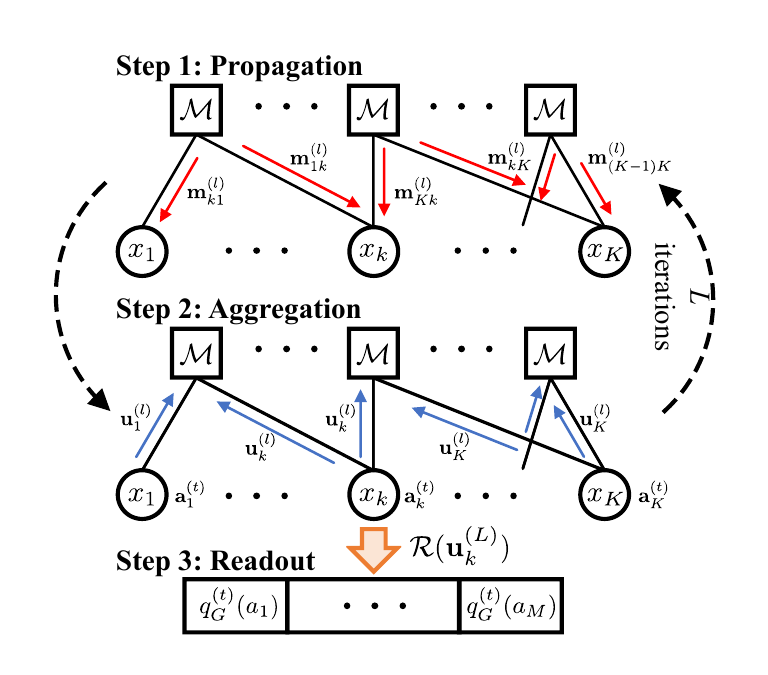}
		\caption{Message passing process of the GNN.}
		\label{fig:gnn}
	\end{figure}
}

% paraphrase 
\subsubsection{Propagation}
The edge 
between any pair of the variable nodes $x_k$ and $x_j$ first concatenates its own feature vector $\mathbf{f}_{jk}$ with the incoming feature vectors $\mathbf{u}_k^{(l-1)}$ and $\mathbf{u}_j^{(l-1)}$. 
Then, the concatenated feature is delivered to a multi-layer perceptron (MLP) for message encoding, and the corresponding output message $\mathbf{m}_{jk}^{(l)}$ is  
\CheckRmv{
	\begin{equation}
		\mathbf{m}_{j k}^{(l)}=\mathcal{M}(\mathbf{u}_{k}^{(l-1)}, \mathbf{u}_{j}^{(l-1)}, \mathbf{f}_{j k}),
		\label{eq:propagation}
	\end{equation}
}
where $\mathcal{M}$ denotes the operation of the MLP, which has two hidden layers of sizes $N_{\rm h1}$ and $N_{\rm h2}$ with the rectifier linear unit (ReLU) as the activation function and an output layer of size $N_{\rm u}$. 
Finally, the message $\mathbf{m}_{jk}^{(l)}$ is sent to the variable node $x_k$, as shown in \figref{fig:gnn}.

\subsubsection{Aggregation}
The aggregation at node $x_k$ is carried out by summing the incoming messages $\mathbf{m}_{jk}^{(l)}$ from the connected nodes $x_j$ and then concatenating the sum with the node attribute $\mathbf{a}_{k}^{(t)}$, given by $\mathbf{m}_{k}^{(l)}=\big[\sum_{j\in [K]\backslash\{k\}} \mathbf{m}_{j k}^{(l)}, \mathbf{a}_{k}^{(t)}\big]$.
Subsequently, a gated recurrent unit (GRU) $\mathcal{U}$ \cite{cho2014learning} is  used to update the node feature vector $\mathbf{u}_k^{(l)}$ by incorporating the concatenated message $\mathbf{m}_{k}^{(l)}$:
\CheckRmv{
	\begin{subequations}
		\begin{align}
			\mathbf{g}_{k}^{(l)}&=\mathcal{U}(\mathbf{g}_{k}^{(l-1)}, \mathbf{m}_{k}^{(l)}), \\
			\mathbf{u}_{k}^{(l)}&=\mathbf{W}_{2} \cdot \mathbf{g}_{k}^{(l)}+\mathbf{b}_{2}, \label{eq:gru_output}
		\end{align}
	\end{subequations}
}
where $\mathbf{g}_{k}^{(l)} \in \mathbb{R}^{N_{\rm h1}}$ and $\mathbf{g}_{k}^{(l-1)}\in \mathbb{R}^{N_{\rm h1}}$ denote the current and previous hidden states, respectively. The single-layer NN with the learnable parameters $\mathbf{W}_2 \in \mathbb{R}^{N_{\rm u} \times N_{\rm h1}}$ and $\mathbf{b}_2 \in \mathbb{R}^{N_{\rm u}}$ in \eqref{eq:gru_output} is used to derive the output feature, which is delivered to the neighboring propagation module for the next MP iteration.

% ok 
\subsubsection{Readout} 
The node feature vectors $\mathbf{u}_{k}^{(L)}$ are sent to the readout module, which contains a MLP followed by the SoftMax function, after $L$ rounds of MP and yields
\CheckRmv{
	\begin{subequations}
		\begin{align}
			\mathbf{z}_k &=\mathcal{R}(\mathbf{u}_{k}^{(L)}), \\
			q_{G}^{(t)}(x_k=a_m)&=\frac{\exp (z_{k, a_m})}{\sum_{a_m \in \mathcal{A}} \exp (z_{k, a_m})}, a_m \in \mathcal{A}, \label{eq:softmax}
		\end{align}
	\end{subequations}
}
where $\mathcal{R}$ is the MLP with two hidden layers of sizes $N_{\rm h1}$ and $N_{\rm h2}$ activated by ReLU and an output layer of size $M$, which is the cardinality of the real-valued constellation $\mathcal{A}$. 
The SoftMax process in \eqref{eq:softmax} maps the unnormalized output $\mathbf{z}_k$ of $\mathcal{R}$ into the 
estimated marginal distribution $q_{G}^{(t)}(x_k)$ of the discrete variable $x_k$ for the $t$-th layer of {\GEP}. 
Moreover, the final GRU hidden state and node feature vector in the current {\GEP} layer are assigned to the next layer as the starting point, that is, $\mathbf{g}_k^{(0)}\leftarrow\mathbf{g}_k^{(L)}\;\text{and}\;\mathbf{u}_k^{(0)}\leftarrow\mathbf{u}_k^{(L)},k\in [K]$.

Notably, the weight parameters of $\mathcal{M}$, $\mathcal{U}$, and $\mathcal{R}$ are shared across different nodes or edges. 
{These parameters can be trained with supervised learning 
to improve EP's estimations characterized only by the mean and variance of a Gaussian function.} 
{This allows the {\GEP} detector to calculate the \post distribution in \eqref{eq:est_post} using the improved estimation $q_{G}^{(t)}(x_k)$ from the GNN instead of $q_{\rm e}^{(t)}(x_k)$.
Hard output $\mathbf{\hat{x}}^{(T)}$ can then be derived based on the \post estimation using \eqref{eq:ep_post}  
in the last layer of the network \cite{kosasih2022graph}.
However, designing the {\GEP}-based turbo receiver is a non-trivial task due to the requirement for extrinsic information, rather than \post information.}

\subsection{Proposed EXT-{\GEP}}

\subsubsection{Motivation}
{\figref{fig:gnn-ep} presents the proposed turbo structure for {\GEP}.
% soft-input 
Similar to the EP-based turbo receiver described in \secref{sec:ep}, the {\GEP} detector in the designed turbo structure integrates priors from the channel decoder.
In \figref{fig:gnn-ep}, the blue dashed lines represent the priors from the channel decoder. 
These priors are used in both the computation of the initial pair, as shown in \eqref{eq:init_pair}, and in the improved \post estimation at the Posterior module as: 
\CheckRmv{
	\begin{equation}
		\hat{p}_{G}^{(t)}(x_k) \propto q^{(t)}_{G}(x_{k}){p_{\rm A1}}(x_k), 
		\label{eq:gep_post}
	\end{equation}
}
which combines the outputs of the GNN with the \prior PDF. 
The estimated \post LLRs $\hat{\mathbf{L}}_{\rm APP}$ can be demapped from the estimated APPs $\hat{p}_{G}^{(T)}(x_k)$ at the last layer of {\GEP}.}

As shown in \figref{fig:gnn-ep}, the stability of the IDD process depends on {\GEP}'s ability to {provide the decoder with} extrinsic LLRs {$\mathbf{L}_{\mathrm{E1}}$} that follow the turbo principle \cite{tuchlerMinimumMeanSquared2002}. 
This means that the extrinsic LLRs should only contain new information and should not count the same information twice.
Recall that the MAP detector produces extrinsic LLRs by subtracting the \prior LLRs {$L_{\mathrm{A1}}(c_j)$} from the \post LLRs:
\CheckRmv{
	\begin{equation} 
		L_{\mathrm{E1}}(c_j) = \log \frac{p\left(c_{j}=1 | \mathbf{y}\right)}{p\left(c_{j}=0 | \mathbf{y}\right)}-L_{\mathrm{A1}}(c_j).
		\label{eq:le1}
	\end{equation}
}
{However, equipping GEPNet with this strategy leads to poor performance, as shown in the simulation results. 
This phenomenon can be attributed to the fact that {\GEP}'s approximation to the MAP detector may not be accurate, 
and directly subtracting the \prior LLRs may not completely eliminate the impact of the priors \cite{huangExtrinsicNeuralNetwork2021}. 
This results in unreliable LLRs that deviate from the desired extrinsic LLRs \cite{papkeImprovedDecodingSOVA1996}.} 
To address this issue, we customize a training scheme for {\GEP} to enable the network to output LLRs that do not couple with the priors.  
{We also utilize a decoder LLR preprocessing strategy to further stabilize the proposed receiver.} 
The details of the resultant turbo receiver scheme 
are revealed in the following subsection. 

\subsubsection{Three-step Training of EXT-{\GEP}} \label{sec:three_step}
\figref{fig:ext-gnn-ep} presents the training procedure for the 
EXT-{\GEP}, which is divided into three steps, which are elaborated as follows.

\CheckRmv{ 
\begin{figure*}[tbp]
	\centering
	\subfigure[Step 1: Train APP-{\GEP}]{
		\includegraphics[width=1.44in]{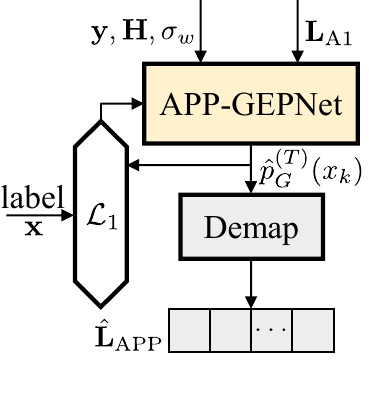}
		\label{fig:step1}
	}
	\subfigure[Step 2: Generate extrinsic training LLRs]{
		\includegraphics[width=3.12in]{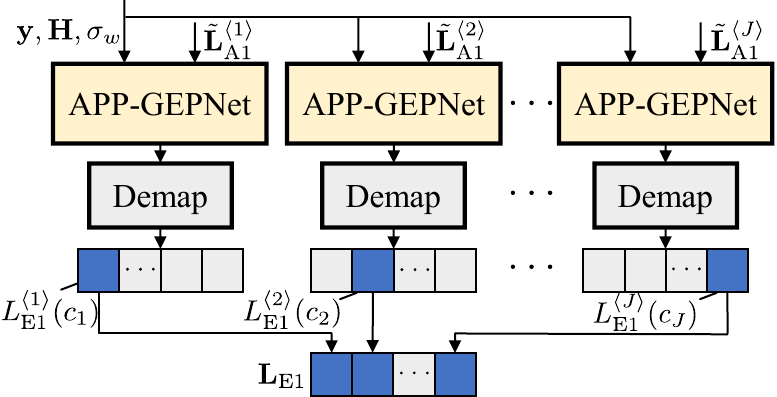}
		\label{fig:step2}
	}
	\subfigure[Step 3: Train EXT-{\GEP}]{
		\includegraphics[width=1.6in]{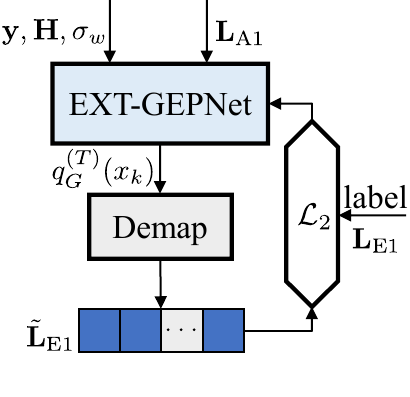}
		\label{fig:step3}
	}
	\caption{Overview of the three-step training process for the EXT-{\GEP}.} 
	\label{fig:ext-gnn-ep}
\end{figure*}
}

\textbf{Step 1: Train an {APP-based} {\GEP}.}
We first train a {\GEP} detector to generate \post LLRs, indicated by APP-{\GEP} in \figref{fig:step1}.  
In this step, a randomly generated pair {$\{\mathbf{x}^{[d]},\mathbf{y}^{[d]},\mathbf{H}^{[d]},\sigma_w^{[d]}\}$} and a bitwise \prior LLR vector {$\mathbf{L}_{\rm A1}^{[d]}$} of size $J=KQ$ form a training sample, where $d$ is the sample index. 
The transmitted symbol vector $\mathbf{x}^{[d]}$ is the label. 
The received signal $\mathbf{y}^{[d]}$, the channel state information (CSI) $\{\mathbf{H}^{[d]},\sigma_w^{[d]}\}$, and the LLR vector $\mathbf{L}_{\rm A1}^{[d]}$ are the input features.
The elements of $\mathbf{L}_{\rm A1}^{[d]}$ are generated according to the  bits $\mathbf{c}^{[d ]}$ demapped from  $\mathbf{x}^{[d ]}$ and are assumed to be Gaussian distributed, {${L}_{\rm A1}(c_j^{[d ]})\sim{\mathcal N}(L;(2c_j^{[d ]}-1){\mu _A},2{\mu _A})$}, where $c_j^{[d ]}$ is the corresponding $j$-th bit. 
{The mean of the Gaussian distribution is $\mu_A = J_A^{-1}(I_A)$, where $J_A(\mu) \triangleq 1-\mathbb{E}_{\mathcal{N}(L;\mu, 2 \mu)}[\log _{2}(1+{\rm e}^{-L})]$ is a monotonically increasing function.}
The parameter $I_A \in [0,1]$ denotes the average mutual information between $\mathbf{L}_{\rm A1}^{[d ]}$ and $\mathbf{c}^{[d ]}$  and characterizes the quality of the \prior LLRs \cite{tenbrinkConvergenceBehaviorIteratively2001}.
This idea is inspired by the simulation of the extrinsic information transfer chart for IDD \cite{tenbrinkConvergenceBehaviorIteratively2001} and allows for an open-loop training without relying on the channel coding scheme.

To supply the network with different \prior LLR distributions in the training stage for the sake of generalization, $I_A$ can be uniformly selected  from [0,1].
However, to avoid the tedious numerical calculations from $I_A$ to $\mu_A$, we set  a look-up table (LUT) between $I_A$ and $\mu_A$ in advance for a predefined set $I_A \in \{0, 0.33, 0.67, 0.78, 0.89, 0.94, 0.99, 1\}$ \cite{sahinDoublyIterativeTurbo2019}. 
This set already reflects the increasing trend of the confidence level, i.e., the absolute value of LLRs from the channel decoder, as the number of TIs increases.
Thus, in our implementation, $I_A$ is randomly selected from the predefined set for each vector $\mathbf{L}_{\rm A1}^{[d ]}$  and converted into the corresponding $\mu_A$ based on the LUT. 
Then, the \prior LLR vector is derived according to the given bits in the training sample, the mean value $\mu_A$, 
and the Gaussian distribution.

{Moreover, we train the network with the cross-entropy (CE) 
loss function according to \cite{kosasih2022graph}}:
\CheckRmv{
	\begin{equation}
		\mathcal{L}_1 = - \frac{1}{D}\sum\limits_{d = 1}^D {\sum\limits_{k = 1}^{{K}} {\sum\limits_{a \in \mathcal{A} } {{\mathbb{I}_{x_k^{[d ]} = a}}} } } \log \big( {\hat{p}_{G}^{(T)}(x_k^{[d ]} = a)} \big),
		% one-hot \times \log(q_G^{(T)}(\mathbf{x}))
		\label{eq:loss1}
	\end{equation}
}
where $D$ is the number of samples in a training batch, $x_k^{[d ]}$ is the $k$-th element of $\mathbf{x}^{[d ]}$, and ${\mathbb{I}_{x_k^{[d ]} = a}}$ is the indicator function that takes the value one if $x_k^{[d ]} = a$ and zero otherwise, hence representing the training label.
By backpropagating the error over this loss, {the estimated APPs $\hat{p}_G^{(T)}(x_k)$ from the APP-{\GEP} 
can gradually approach the true APPs of the transmitted symbols. Finally, $\hat{p}_G^{(T)}(x_k)$ can be demapped as the estimated \post LLRs {$\hat{\mathbf{L}}_{\rm APP}$}.} 

\textbf{Step 2: Generate extrinsic training LLRs from APP-{\GEP}.}
{The extrinsic LLR in \eqref{eq:le1} can be rewritten as \cite{tuchlerMinimumMeanSquared2002}:
\CheckRmv{
	\begin{align}
		L_{\mathrm{E1}}(c_j) &= \log \frac{\sum_{\forall \mathbf{c}: c_{j}=1} p(\mathbf{y} | \mathbf{c}) p_{\rm A1}(\mathbf{c})}{\sum_{\forall \mathbf{c}: c_{j}=0} p(\mathbf{y} | \mathbf{c}) p_{\rm A1}(\mathbf{c})}-L_{\mathrm{A1}}(c_j) \nonumber \\
		&=\log \frac{\sum_{\forall \mathbf{c}: c_{j}=1} p(\mathbf{y} | \mathbf{c}) \prod_{\forall i:i\neq j} p_{\rm A1}(c_i)}{\sum_{\forall \mathbf{c}: c_{j}=0} p(\mathbf{y} | \mathbf{c})\prod_{\forall i:i\neq j} p_{\rm A1}(c_i)}.
		\label{eq:le1_split}%
	\end{align}}
This equation demonstrates that the extrinsic LLR $L_{\mathrm{E1}}(c_j)$ is a function of the channel information and the priors $L_{\mathrm{A1}}(c_i), \forall i\neq j$ and should be independent of $L_{\mathrm{A1}}(c_j)$ \cite{tuchlerMinimumMeanSquared2002}.}
Special inputs are designed for the APP-{\GEP} obtained in Step 1 so that the network can generate extrinsic LLRs that satisfy this criterion. 
% ok 
First, a total of $J$ modified \prior LLR vectors {$\{\tilde{\mathbf{L}}_{\rm A1}^{\langle j\rangle}\}_{j=1}^{J}$} are derived by setting the $j$-th ($j \in [J]$) element of the original \prior LLR vector $\mathbf{L}_{\rm A1}$ to zero, where the sample index $d$ is omitted for ease of notation. 
Thus, the $i$-th element of the $j$-th modified vector $\tilde{\mathbf{L}}_{\rm A1}^{\langle j\rangle}$ is defined as 
\CheckRmv{
	\begin{equation} 
		\tilde L_{\rm A1}^{\langle j\rangle}(c_i) = \left\{ {\begin{array}{*{20}{l}}
		{L_{{\rm{A1}}}(c_i)}&{{\rm{ if }}\;i \ne j}\\
		0&{{\rm{ if }}\;i = j}
		\end{array}} \right.,i\in [J].
	\end{equation}
}

Next, the modified \prior LLR vectors {$\{\tilde{\mathbf{L}}_{\rm A1}^{\langle j\rangle}\}_{j=1}^{J}$}, along with {the same features {$\{\mathbf{y},\mathbf{H},\sigma_w\}$} corresponding to $\mathbf{L}_{\rm A1}$ from Step 1,} are used as the inputs for the APP-{\GEP} to derive the extrinsic LLRs via $J$ parallel inferences, as shown in \figref{fig:step2}.
Specifically, the inputs are $\tilde{\mathbf{L}}_{\rm A1}^{\langle j\rangle}$ and $\{\mathbf{y},\mathbf{H},\sigma_w\}$ when we target at the $j$-th modified vector.
The LLR $L_{\rm E1}^{\langle j\rangle}(c_j)$ from the corresponding output vector is hence independent of the initial prior ${L_{{\rm{A1}}}}(c_j)$ because $\tilde L_{{\rm{A1}}}^{\langle j\rangle}(c_j)$ is assigned to zero. 
{Therefore, $L_{\rm E1}^{\langle j\rangle}(c_j)$ is exempted from coupling with the corresponding bit prior and satisfies the criterion for extrinsic LLR given in \cite{tuchlerMinimumMeanSquared2002}.}  
{Finally, $L_{\rm E1}^{\langle j\rangle}(c_j)$ is collected individually from each output vector  according to \figref{fig:step2} to form {$\mathbf{L}_{\rm E1}=\{L_{\rm E1}^{\langle j\rangle}(c_j)\}_{j=1}^{J}$}.
These LLRs are then used as the training labels for EXT-{\GEP} in the next step. }

\textbf{Step 3: Train the final EXT-{\GEP}.}
In this step, we train the final EXT-{\GEP} to learn the mapping from the input features to the {extrinsic LLRs} derived in Step 2. 
Notably, the EXT-{\GEP} shares the same structure with the APP-{\GEP} trained in Step 1, given by \figref{fig:gnn-ep}, whereas {the difference lies in two aspects: First, for EXT-{\GEP}, we use the outputs of the GNN $q_G^{(T)}(x_k)$ instead of the estimated APPs $\hat{p}_G^{(T)}(x_k)$ to derive the output LLRs $\tilde{\mathbf{L}}_{\rm E1}$. 
This avoids coupling with the priors.
Second, the objective of the training and the resultant model weights are different.}  
Specifically, the loss function for training the EXT-{\GEP} is
\CheckRmv{
	\begin{equation} 
		\mathcal{L}_2 =  \frac{1}{D}\sum\limits_{d = 1}^D { {\sum\limits_{j = 1}^J {{l_{\rm CE}}(c_{{\rm e},j}^{[d ]},\tilde c_{{\rm e},j}^{[d ]})} } }, 
		\label{eq:loss_ext}
	\end{equation}
}
where {$c_{{\rm e},j}^{[d ]}$} represents a soft bit mapped from the {extrinsic LLR sample} {$L_{\rm E1}^{[d ]}(c_j)$} of Step 2, and {$\tilde c_{{\rm e},j}^{[d ]}$} denotes the counterpart equivalent of the output LLR {$\tilde L_{\rm E1}^{[d ]}(c_j)$} of the target EXT-{\GEP}: 
\CheckRmv{
	\begin{equation*} 
		c_{{\rm e},j}^{[d ]}=\frac{1}{1+\exp(-L_{\rm E1}^{[d ]}(c_j))},~~ \tilde c_{{\rm e},j}^{[d ]}=\frac{1}{1+\exp(-\tilde L_{\rm E1}^{[d ]}(c_j))}.
	\end{equation*}
}
Furthermore, $l_{\rm CE}$ in \eqref{eq:loss_ext} is given by
\CheckRmv{
	\begin{equation}
		{l_{\rm CE}}({c},{\tilde c}) =  - \big({c}\log ({\tilde c}) + (1 - {c})\log (1 - {\tilde c})\big).
	\end{equation}
}
Therefore, the output LLRs $\tilde{\mathbf{L}}_{\rm E1}$ of the EXT-{\GEP} 
can approximate the desired output ${\mathbf{L}}_{\rm E1}$ from Step 2 by error backpropagation over the loss $\mathcal{L}_2$, as shown in \figref{fig:step3}.  

{The three-step training scheme of the EXT-GEPNet is summarized in \algref{alg:training}.\footnote{Note that the randomly generated data pairs $\{\mathbf{y},\mathbf{H},\sigma_w, \mathbf{L}_{\rm A1}\}$ in Step 1 are used throughout the three-step training scheme. No additional data pairs are generated specifically for training the EXT-GEPNet.}}

\setlength{\algomargin}{0em} % indent space inside the algorithm (\leftskip + \parindent)
\SetAlCapHSkip{0em} % Set algorithm horizontal skip
\CheckRmv{
	\begin{algorithm}[!t]
	\caption{Three-step training of EXT-GEPNet}
	\label{alg:training}
	{\small
	\begingroup
	\SetInd{0.5em}{0.5em}
	%\addtolength{\jot}{0.3em}
	\KwIn{Initial network parameters; dataset sizes $D_1$ and $D_2$; $I_A \in \{0, 0.33, 0.67, 0.78, 0.89, 0.94, 0.99, 1\}$ and 
	$\mu_A = J_A^{-1}(I_A)$.} 
	\KwOut{The trained network parameters.}
	\vspace{0.1cm}
	\hrule
	\vspace{0.1cm}

	\textbf{Train an APP-based GEPNet:} \\
	\quad Randomly generate a training dataset $\mathcal{D}_1$ with $D_1$ samples,
	\setlength\abovedisplayskip{0pt}
	\setlength\belowdisplayskip{0pt}
	\begin{align*}
		\text{each sample: } & \{\mathbf{x},\mathbf{y},\mathbf{H},\sigma_w,  \mathbf{L}_{\rm A1}\},\; \text{label: } \mathbf{x},\\ 
		\textit{a priori } \text{LLR: } &  {L}_{\rm A1}(c_j)  \sim{\mathcal N}(L;(2c_j-1){\mu _A},2{\mu _A}); 
	\end{align*}

	\quad Train the GEPNet using $\mathcal{D}_1$ to minimize $\mathcal{L}_1$ in (21); \\
	\quad Derive the trained network parameters for an APP-GEPNet.\\

	\vspace{0.1cm}
	\hrule
	\vspace{0.1cm}

	\textbf{Generate extrinsic training LLRs from APP-GEPNet:} \\
	\For{$d=1$ \KwTo $D_2$}{
		Derive $J$ modified \prior LLR vectors $\{\tilde{\mathbf{L}}_{\rm A1}^{\langle j\rangle}\}_{j=1}^{J}$ %for each sample in Step 1 
		via (23);\\
		Perform $J$ parallel inferences via the APP-GEPNet;\\
		Collect the output extrinsic LLRs  
		to form $\mathbf{L}_{\rm E1}=\{L_{\rm E1}^{\langle j\rangle}(c_j)\}_{j=1}^{J}$. 
	}

	\vspace{0.1cm}
	\hrule
	\vspace{0.1cm}

	\textbf{Train the final EXT-GEPNet:}\\
	\quad Training dataset $\mathcal{D}_2$ with $D_2$ samples,
	\begin{equation*}
		\text{each sample: }  \{\mathbf{y},\mathbf{H},\sigma_w,  \mathbf{L}_{\rm A1},\mathbf{L}_{\rm E1}\},\; \text{label: } \mathbf{L}_{\rm E1}; % for the entire sample/ each sample
	\end{equation*}

	\quad Train the GEPNet using $\mathcal{D}_2$ to minimize $\mathcal{L}_2$ in (24); \\
	\quad Derive the trained network parameters for an EXT-GEPNet.\\
	\endgroup
	}
	\end{algorithm}
}

\begin{remark}
Once the EXT-{\GEP} is trained, it can be deployed in the turbo structure of \figref{fig:gnn-ep} and used for different TIs with the same model parameters, constructing the EXT-{\GEP}-based turbo receiver. 
As the proposed method is an approximation to the MAP receiver, 
overestimation of the reliability can happen during the IDD procedure. 
This overestimation can result in instability in the detection process as the mean value of the closed-loop LLRs from the channel decoder would increase over iterations.
We observed experimentally that scaling the decoder LLRs into the range that matches the range of the \prior LLRs $\mathbf{L}_{\rm A1}$ used in the training stage effectively overcomes the instability issues.
Therefore, we utilize an adaptive LLR scaling method \cite{huangExtrinsicNeuralNetwork2021} involving three steps: 
First, we examine the training LLRs $ \mathbf{L}_{\rm A1}$ to determine the range $[-r,r]$ that $ \mathbf{L}_{\rm A1}$ fall into with a probability of $p_r\approx 1$.
Second, we search for the maximum absolute value $r_{\iota}$ of the decoder LLRs in the current codeword at the end of each TI.
Finally, we scale the decoder LLRs by $r_{\iota}/r$ if $r_{\iota} > r$; otherwise, we keep the LLRs unchanged. 
As a result, the decoder LLRs are controlled in an appropriate range.
In this work, we set $p_r$ as 0.97 and find that the decoder LLR preprocessing method further stabilizes the IDD procedure. 
\end{remark}

\subsection{Edge Pruning to Reduce Complexity}  
% Motivation ok
The most computationally demanding operation of the GNN 
lies in the message propagation step through all the edges of the graph shown in \figref{fig:gnn}.
This step involves the execution of the edge MLPs, i.e., the function $\mathcal{M}$ in \eqref{eq:propagation}, for $K(K-1)$ times because the MRF defined in Section \ref{sec:GNN-EP} is FC{, i.e., with each pair of nodes connected by an edge}. 
However, considerable redundant connections are observed in this FC graph \cite{scotti2020graph}.

% method
%% ok
We propose an edge pruning scheme based on the covariance matrix $\boldsymbol{\Sigma}$ calculated by the LMMSE module of EP in \eqref{eq:sigma} to simplify the MP of the GNN while maintaining competitive performance.\footnote{The edge pruning is implemented in each layer of the network, and thus {the index ${(\iota,t)}$ for $\boldsymbol{\Sigma}$} is omitted.}
%% ok 
$\boldsymbol{\Sigma}$ is used as an indicator of the correlation weights between the neighboring nodes in the FC graph 
because the element $\Sigma_{ij}$ ($i,j\in [K]$) reflects the covariance of variable nodes $x_i$ and $x_j$.
%% ok
$\Sigma_{ij}$ can be normalized to derive the correlation coefficient as {$\rho_{ij}=\frac{\Sigma _{ij}}{\sqrt{{\Sigma _{ii}}{\Sigma _{jj}}}}\in [0,1]$}. 
%% ok
A large correlation coefficient reflects the high structural dependency between the two connected variables.
Hence, more attention should be paid to the edge between them.
%% ok
On the other hand, a small $\rho_{ij}$ suggests the approximate independence of the two variables, where less information is required to exchange 
between them to recover the transmitted symbols. 
%% ok
Hence, the proposed scheme prunes those edges with correlation coefficients $\rho_{ij}$ 
that meet the following criterion:   
\CheckRmv{
	\begin{equation}
		{\rho_{ij} ^2} < \alpha  \cdot \frac{1}{{K - 1}}\sum\limits_{k = 1,k \ne j}^{K} {{\rho _{kj}}^2}, 
		\label{eq:pruning}
	\end{equation}
}
where $\alpha$ is a positive factor that controls the pruning threshold, 
{and the edge pruning version of the {\GEP}-based method is indicated by this factor hereafter.}
This strategy means that for a specific node $x_j$, only the incoming edges with correlation coefficients $\rho_{ij}$ larger than the average number of all connected edges (multiply by the pruning factor $\alpha$) are retained, and a large $\alpha$ intuitively results in a significant proportion of removed edges. 
% effect ok
Therefore, edges with small correlation weights, i.e., low contributions to the inference of the target probability, are pruned.
This process reduces the computational  cost of GNN and saves computational resources for tuning the vital parts of the network.   
Notably, edges with low correlation weights exert minimal impact on the overall performance.
Thus, the sparsely connected network after pruning can control the performance loss within the acceptable range or be even more generalizable, which is consistent with the findings in \cite{buchbergerPruningQuantizingNeural2021}.
{
\begin{remark}
	Note that while the proposed edge pruning strategy shows potential in improving performance by mitigating overfitting, 
	its underlying mechanism differs from conventional techniques aimed at alleviating overfitting,
	such as early stopping based on a separate validation dataset \cite{goodfellowDeepLearning2016}.  
	Specifically, the early stopping technique prevents overfitting by halting training when the validation loss begins to increase.
	In contrast, the proposed edge pruning focuses on reducing model complexity, thus enhancing generalization through the elimination of unnecessary connections.
	This approach makes the network more efficient during the inference stage, which cannot be achieved through early stopping alone.
\end{remark}}

%%%%%%%%%%%%%%%%%%%%%%%%%%%%%%%%%%%%%%%%%%%%%%%%%%%%%%%%%%
\section{Simulation Results}    \label{sec:simu_results}
The numerical results of the proposed schemes are presented in this section. First, the parameter settings are introduced. 
Second, the performance of the proposed schemes is evaluated under  uncoded and coded MIMO systems.  
Finally, the computational complexity is analyzed.   

\subsection{Parameter Settings}  

The simulated signal-to-noise ratio (SNR) of the system is defined as $\text{SNR}=\frac{\mathbb{E}[\|\mathbf{Hx}\|^2]}{\mathbb{E}[\|\mathbf{w}\|^2]}$.
The i.i.d. Rayleigh and spatially correlated channels are used in the simulation. 
The Rayleigh MIMO channel $\mathbf{H}$ has elements drawn from the Gaussian distribution $\mathcal{N}(h_{ij};0,1/N)$, where $h_{ij}$ is the $(i,j)$-th element of $\mathbf{H}$.
The spatially correlated channel is characterized by the Kronecker model as
${\mathbf{H}} = {\mathbf{R}}_{\rm r}^{1/2}{\mathbf{UR}}_{\rm t}^{1/2}$,
where $\mathbf{U}$ is the i.i.d. Rayleigh channel matrix. $\mathbf{R}_{\rm r}$ and $\mathbf{R}_{\rm t}$, which have exponential elements that correspond to the same spatial correlation coefficient $\varrho$ \cite{loyka_channel_2001}, represent the correlation matrices at the receiver and  transmitter, respectively. 

To evaluate the error rate performance, the maximum number of transmitted bits is set as $5\times 10^7$.
For uncoded systems, the symbol error rate (SER) is used as the performance metric, while for the turbo receiver, the coded bit error rate (BER) and {word error rate (WER) with $N_{\rm b}$ as the word length} are used.
% channel coding
For the coded MIMO systems, convolutional codes (CCs) and turbo codes are selected as the channel coding scheme. 
The CCs have generator polynomial $[133_{o}\;171_o]$. 
{Random interleaving and} 
two code rates are adopted: one is code rate $R_{\rm c}=1/2$ with word length $N_{\rm b}=128$, and the other is $R_{\rm c}=5/6$ with $N_{\rm b}=800$.
The word length is chosen so that the CCs can achieve effective error correction.
{The turbo codes have a code rate of $R_{\rm c}=1/2$ and word length $N_{\rm b}=1952$, and the interleaver follows the 3rd Generation Partnership Project Release 17 specification \cite{3gpp36212r14-2021}. 
The channel decoder utilizes the BCJR algorithm \cite{bahl1974optimal} with 10 inner iterations.}

The hyperparameters of the GNN are set as $N_{\rm h1}=64,N_{\rm h2}=32,N_{\rm u}=8$, and $L=2$.
% dataset ok
For the {\GEP} detector in uncoded systems, the network is trained with the loss function $\mathcal{L}_1$ in \eqref{eq:loss1} but without the \prior training LLRs.
The training and validation sets contain 6,400,000 and 6,000 samples, respectively.
{Furthermore, during the three-step training scheme as described in \secref{sec:three_step}}, training and validation sets for APP-{\GEP} in Step 1 have the same sizes as those for {\GEP} in uncoded systems.
Then, a total of 76,800 extrinsic LLR {vectors $\mathbf{L}_{\rm E1}$} is generated in Step 2 to train EXT-{\GEP} in Step 3. 
The objective is to minimize the loss function $\mathcal{L}_2$ as described in \eqref{eq:loss_ext}. 
{Although this dataset is relatively small, it significantly reduces the computational cost associated with generating the extrinsic training LLRs in Step 2. Through experiments, we have observed that this small dataset is sufficient for training the EXT-GEPNet to effectively learn the mapping from the input features to the extrinsic LLRs, resulting in excellent performance.}
All the considered networks are trained for 5,000 epochs with a batch size of 128.

{In our experiment, we utilize the Glorot normal initializer \cite{glorotUnderstandingDifficultyTraining2010} to initialize the weights of the network. Notably, the APP-GEPNet trained in Step 1 can also serve as an initialization model for the EXT-GEPNet in Step 3. This initialization strategy can help accelerate the convergence speed of the training procedure.} 
{The SNR during training is set as a specific point $\text{SNR}_{\text{train}}$.}
Moreover, the optimizer is selected as Adam with a learning rate of 0.001. 
Unless otherwise specified, we train and test the network under the same
modulation scheme, antenna configurations, and channel model. 
{For uncoded MIMO systems, we choose the damping factor for EP and {\GEP}-based detectors as $\beta=0.2$, as suggested in \cite{cespedesExpectationPropagationDetection2014}.
For coded MIMO systems, we follow the configurations of EP and double EP (DEP), which introduces EP in both the estimation of the posterior and the processing of the channel decoder's feedback to accelerate convergence, as proposed in \cite{santosSelfTurboIterations2019} and \cite{murillo-fuentesLowComplexityDoubleEPBased2021}, respectively.
{The number of layers in all the evaluated iterative detectors is set as $T=5$.}}

\subsection{Performance Analysis of Uncoded MIMO Detectors} \label{sec:uncoded}

\CheckRmv{
\begin{figure}[tbp]
	\centering
	\subfigure[\Times{4}{4} MIMO with \SNRTRAIN{=}{22}]{
		\includegraphics[width=3in]{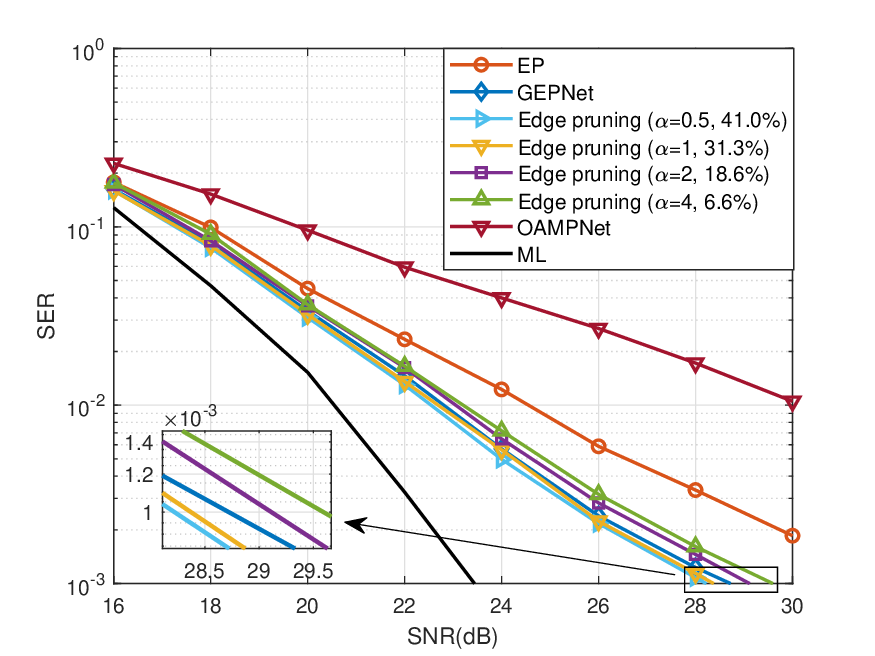}
		\label{fig:ser_4mimo_16qam}
	}
	\subfigure[\Times{16}{16} MIMO with \SNRTRAIN{=}{20}]{
		\includegraphics[width=3in]{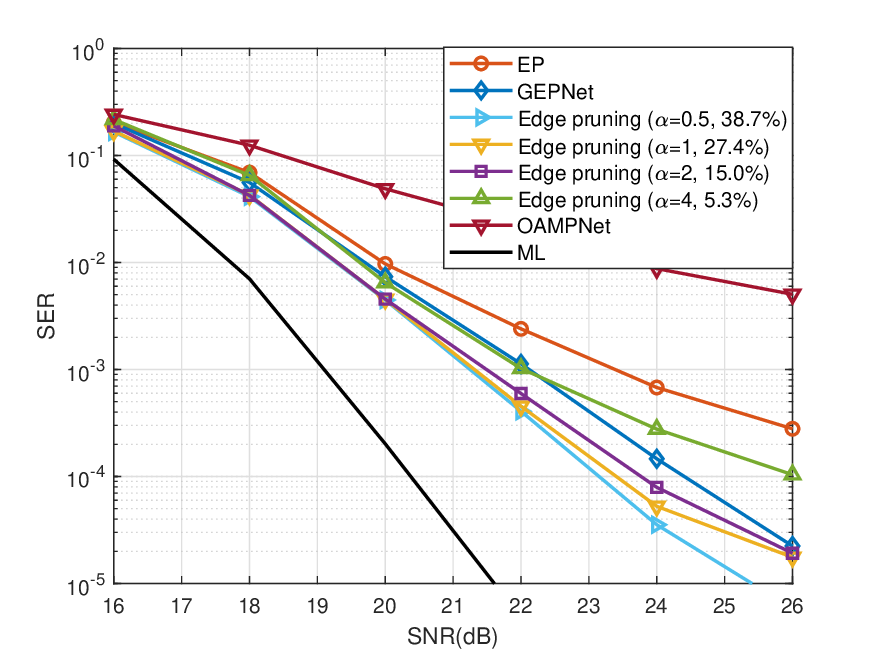}
		\label{fig:ser_16mimo_16qam}
	}
	\caption{SER performance for 16-QAM {under Rayleigh MIMO channels}.}
	\label{fig:ser_16qam}
\end{figure}
}

{We compare the performance of the uncoded MIMO detectors under i.i.d. Rayleigh channels.
\figref{fig:ser_16qam} provides the SER performance under 16-QAM modulation with \Times{4}{4} and \Times{16}{16} MIMO configurations. 
The {\GEP} detector and the edge pruning versions are compared with the conventional EP \cite{cespedesExpectationPropagationDetection2014}, the model-driven DL-based OAMPNet \cite{heModelDrivenDeepLearning2020}, and the optimal ML detectors.}  

{\figref{fig:ser_16qam} reveals that the gain of {\GEP} over EP and OAMPNet is remarkable.}
Moreover, the effect of edge pruning with different pruning factors is demonstrated.
In the comparison, the pruning factor $\alpha$ is set as 0.5, 1, 2, and 4, respectively. 
The number after $\alpha$ of each SER curve of the edge pruning versions in the figure corresponds to the percentage of remaining edges after pruning.
The figure demonstrates that the edge pruning versions with $\alpha=0.5$ and $1$ 
{outperform the original FC model}, thereby revealing the gain brought by appropriately pruning the redundant edges with low correlation weights.
However, the performance of the pruned {\GEP} degrades significantly when $\alpha$ further increases, falling behind the original FC model at $\alpha=2$ and $\alpha=4$ for the \Times{4}{4} and \Times{16}{16} systems, respectively. 
This result is because some dominant edges are improperly removed when $\alpha$ is set too high. 
However, our scheme with $\alpha=4$ (over 90\% edges pruned) still outperforms EP by a large margin because the network is fine-tuned on the basis of EP, thereby indicating its outstanding ability in balancing performance and complexity.

\subsection{Performance Analysis of Coded MIMO Turbo Receiver} \label{sec:coded}
We consider a \Times{4}{4} coded MIMO system with 16-QAM modulation unless noted otherwise. 
First, we separately 
investigate the effect of the three-step training scheme and edge pruning on the proposed turbo receiver under CCs and Rayleigh MIMO channels. 
{Second, we demonstrate the generalization ability of the proposed receiver by evaluating the performance under various mismatches, including channel coding scheme, channel model, SNR, and antenna configuration mismatches.
Finally, we present the robustness of the proposed method against imperfect CSI.}

\subsubsection{Impact of the Three-step Training Scheme} 
\CheckRmv{
\begin{figure}[t]
	\centering
	\subfigure[$R_{\rm c}=1/2$, $N_{\rm b}=128$, and \SNRTRAIN{=}{13}]{
		\includegraphics[width=3in]{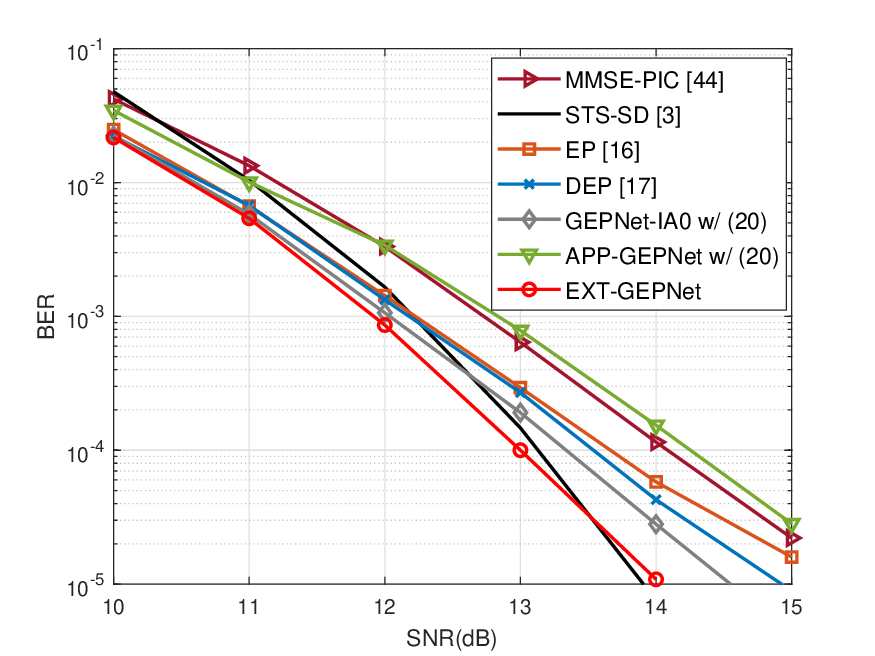}
		\label{fig:ber_4mimo_16qam_cc12}
	}
	\subfigure[$R_{\rm c}=5/6$, $N_{\rm b}=800$, and \SNRTRAIN{=}{18}]{
		\includegraphics[width=3in]{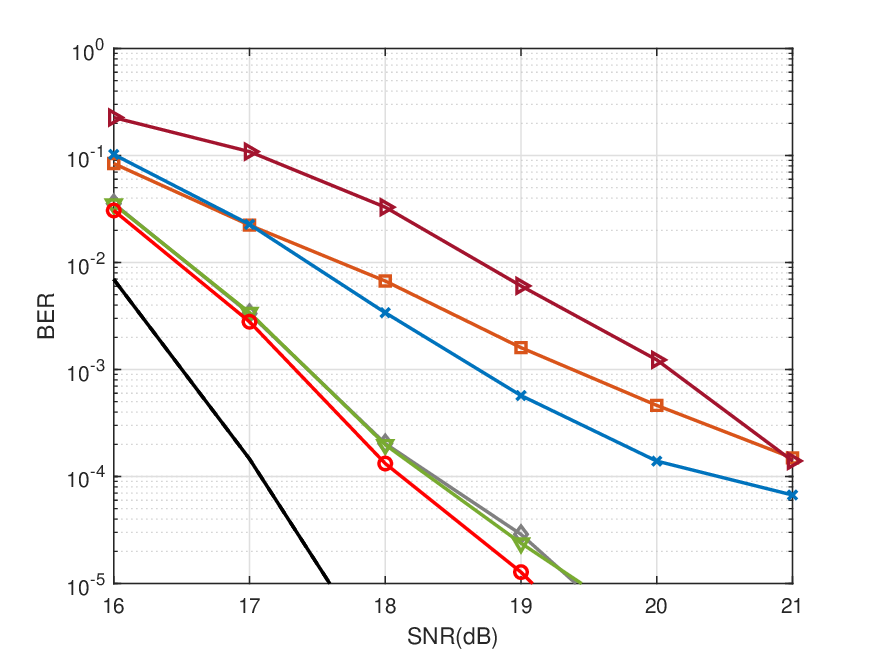}
		\label{fig:ber_4mimo_16qam_cc56}
	}
	\caption{BER performance ($I=2$) with CCs for a \Times{4}{4} Rayleigh MIMO channel with 16-QAM.  
	}
	\label{fig:ber_4mimo_16qam_cc}
\end{figure}
}
{\figref{fig:ber_4mimo_16qam_cc} presents the BER performance comparison 
under CCs and $I=2$.} 
Two {\GEP} baselines are set to validate the proposed training scheme: 
The first is the APP-{\GEP}, derived in Step 1 of the training scheme, and 
the second is the {\GEP} detector trained in the uncoded systems and deployed in the {proposed turbo scheme as shown in \figref{fig:gnn-ep}}. 
This baseline has the same loss function {and \post outputs} as the APP-{\GEP} but with the \prior training LLRs set to zero, i.e., $I_A=0$ for Step 1 of the training scheme, denoted as {\GEP}-IA0 hereafter.
{Both of these baselines are equipped with \eqref{eq:le1} to subtract the \prior information from the outputs {$\hat{\mathbf{L}}_{\rm APP}$}, denoted by ``w/ \eqref{eq:le1}'' in the figures.}{\footnote{{We also conducted tests on the baselines that utilize  \eqref{eq:ext_llr} to demap the output PDF of the GNN module as the extrinsic LLRs. However, we did not observe any performance improvement compared to using \eqref{eq:le1}. Therefore, in order to maintain consistency with the proposed scheme in terms of removing priors and deriving extrinsic LLRs, we equipped the baselines with \eqref{eq:le1} in our simulations.}}}

{\figref{fig:ber_4mimo_16qam_cc12} shows the comparison under code rate $R_{\rm c}=1/2$ and word length $N_{\rm b}=128$.  
APP-{\GEP} performs poorly, which reflects that simply using \eqref{eq:le1} cannot completely remove the correlation of the priors at the outputs to generate reliable extrinsic LLRs.
{\GEP}-IA0 outperforms APP-{\GEP}, which can be attributed to the limited \prior information coupled {during training}, and the correlation problem is less severe. 
However, this baseline cannot effectively manage the diverse \prior information provided by the decoder as it is trained with $I_A =0$. 

EXT-{\GEP} has significant performance gains over the two baselines because the proposed method not only resolves the information coupling problem systematically to generate appropriate extrinsic information, but also fully utilizes the \prior information during training to generalize to different \prior LLR distributions.}
Moreover, the EXT-{\GEP}-based turbo receiver is also compared with
other turbo approaches equipped with the MMSE-based parallel interference cancellation (MMSE-PIC) \cite{witzkeIterativeDetectionMIMO2002}, {EP \cite{santosSelfTurboIterations2019}, {DEP} \cite{murillo-fuentesLowComplexityDoubleEPBased2021}}, and STS-SD \cite{studerSoftInputSoft2010}. 
% ok
The figure shows that EXT-{\GEP} 
outperforms {MMSE-PIC, EP, and DEP by approximately 1 dB} at the BER of $10^{-5}$ and reveals equivalent or even superior performance to the computationally expensive STS-SD at all tested SNRs.

In another aspect, the results from \figref{fig:ber_4mimo_16qam_cc56} reveal that APP-{\GEP} achieves a competitive performance under CCs with a high code rate ($R_{\rm c}=5/6$).
{This can be attributed to the low redundancy channel code resulting in less \prior information being coupled in APP-{\GEP}'s outputs, and the correlation issue is less pronounced compared to that at a code rate of 1/2.}  
However, EXT-{\GEP} still {has superiority over the two {\GEP}-based baselines}, outperforms {MMSE-PIC, EP, and DEP}, and is only inferior to  the high-complexity STS-SD.

\CheckRmv{
	\begin{figure}[t]
		\centering
		\subfigure[$I_A=0.8$]{
			\includegraphics[width=3in]{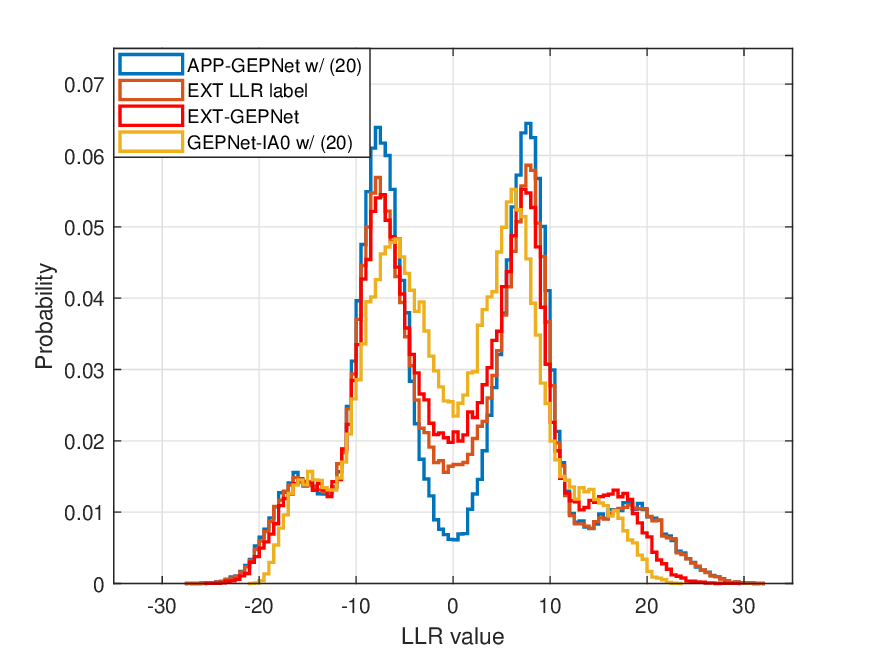}
			\label{fig:ia08}
		}
		\subfigure[$I_A=0.2$]{
			\includegraphics[width=3in]{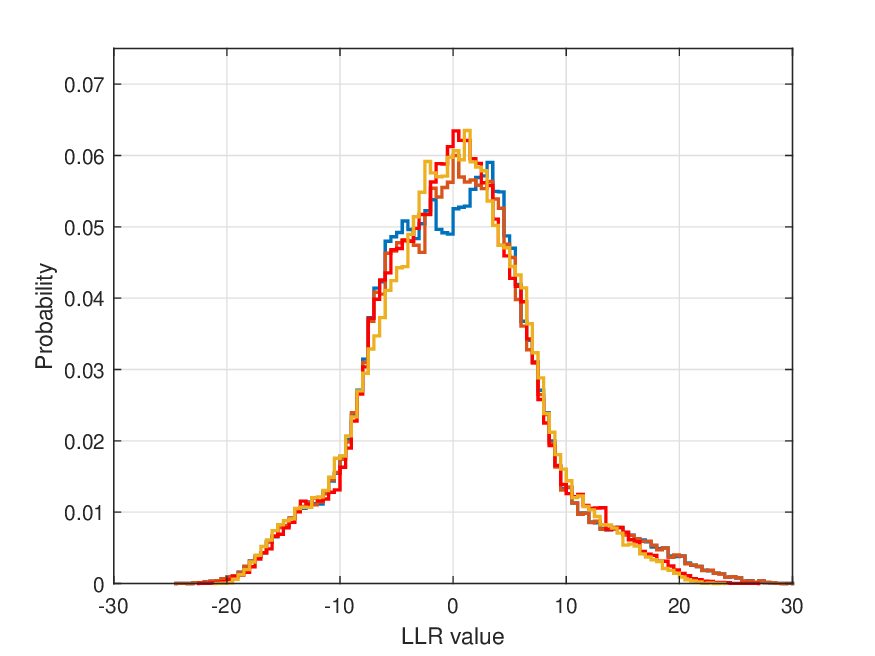}
			\label{fig:ia02}
		}
		\caption{{Output LLR distributions of the detectors for a \Times{4}{4} Rayleigh MIMO channel with 16-QAM and \SNR{=}{13}. 
		}}
		\label{fig:llr_distribution}
	\end{figure}
}
\figref{fig:llr_distribution} provides an intuitive interpretation for the performance in \figref{fig:ber_4mimo_16qam_cc} by analyzing the output LLR distributions of the detector during the training scheme. 
\figref{fig:ia08} shows the results under Gaussian \prior LLRs with $I_A=0.8$, emulating the LLRs from the decoder under CCs with $R_{\rm c}=1/2$. 
Two steep peaks can be found in the output LLRs of APP-{\GEP} even after the explicit subtraction of priors, which likely contain over-optimistic LLR estimations{\footnote{{%that is, incorrect LLR values large in magnitudes with over-confidence.
Over-optimistic LLR estimations refer to the phenomenon where the soft outputs of the detector overestimate the reliability, leading to incorrect LLR estimates with erroneously assigned large magnitudes \cite{papkeImprovedDecodingSOVA1996,vogtImprovingMaxlogMAPTurbo2000}.}}} with residual priors that lead to the poor performance in \figref{fig:ber_4mimo_16qam_cc12}. 

Step 2 of the training scheme, with output LLRs denoted by ``EXT LLR label'' in \figref{fig:ia08}, effectively reduces the magnitudes of the two LLR peaks and alleviates the overestimation issue.  
Moreover, the output LLRs of EXT-{\GEP} 
closely track the LLR labels derived in Step 2, approaching desirable extrinsic LLRs. 
We also provide the output LLR distributions of {\GEP}-IA0 with \eqref{eq:le1} 
as compared to the proposed scheme.
The figure reveals that {\GEP}-IA0 is less capable of distinguishing bits 0 and 1 than EXT-{\GEP}, producing more LLRs with small magnitudes near 0.

\figref{fig:ia02} shows that the difference between the compared distributions is less perceivable when $I_A=0.2$, i.e., limited priors are provided, as compared to $I_A=0.8$, clarifying the competitive performance of APP-{\GEP} under $R_{\rm c}=5/6$.

{\figref{fig:convergence} shows the BER convergence performance across TIs.
We consider a \Times{4}{4} Rayleigh MIMO system with CCs ($R_{\rm c}=1/2$, $N_{\rm b}=128$), 16-QAM, and \SNR{=}{12}. 
The figure shows that EXT-GEPNet achieves the fastest convergence speed 
among the compared schemes. 
The comparison between EXT-GEPNet and GEPNet-IA0 also confirms that EXT-GEPNet better adapts to different TIs than the original GEPNet without the proposed training scheme.
\CheckRmv{
	\begin{figure}[t]
		\setlength{\abovecaptionskip}{-0.2cm}
		\setlength{\belowcaptionskip}{-0.0cm}
		\centering
		\includegraphics[width=3in]{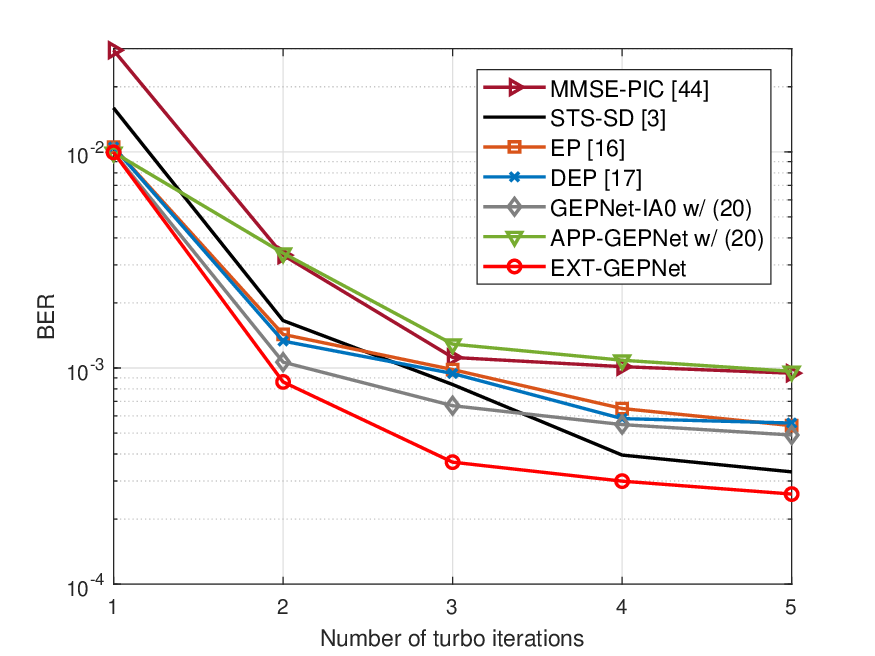}
		\caption{{BER performance across turbo iterations with CCs ($R_{\rm c}=1/2$, $N_{\rm b}=128$) for a \Times{4}{4} Rayleigh MIMO channel with 16-QAM and \SNR{=}{12}.}}
		\label{fig:convergence}
	\end{figure}
}
}

\subsubsection{Impact of Edge Pruning}

\CheckRmv{
	\begin{table}[t]
		\centering
		\caption{Impact of Edge Pruning on the WER Performance of EXT-{\GEP}-based Turbo Receivers ($I=2$)}
		\begin{tabular}{c|c|c}
		\toprule
		EXT-{\GEP}& $R_{\rm c}=1/2$, \SNR{=}{13}  &  $R_{\rm c}=5/6$, \SNR{=}{18} \\  
		\midrule
		${\alpha= 0}$    & 8.81e-4 (100\%)         & 4.22e-3 (100\%)      \\
		$\alpha=0.5$     & \textbf{8.31e-4} (42.9\%)      & \textbf{3.79e-3} (36.1\%)   \\
		$\alpha=1$       & 8.60e-4 (30.4\%)      & 5.81e-3 (28.0\%)   \\
		$\alpha=2$       & 8.65e-4 (18.5\%)      & 6.18e-3 (18.3\%)   \\
		$\alpha=4$       & 1.02e-3 (6.8\%)      & 7.96e-3 (8.2\%)   \\ 
		\bottomrule
		\multicolumn{3}{l}{Note: (\textasciitilde) represents the proportion of remaining edges after pruned.} 
		\end{tabular}
		\label{tab:impact_ed}
	\end{table}
}
In the above analysis of the turbo receiver, all EXT-{\GEP}s use the FC model without edge pruning.
In this subsection, we analyze the impact of the edge pruning method on the balance of performance and complexity.
\tabref{tab:impact_ed} shows the WER performance of EXT-{\GEP}-based turbo receivers with different edge pruning factors $\alpha$ at $I=2$ and representative SNRs, i.e., \SNR{=}{13} for $R_{\rm c}=1/2$ and \SNR{=}{18} for $R_{\rm c}=5/6$. 

We first focus on the results with $R_{\rm c}=1/2$ and \SNR{=}{13}. 
Similar to the results in uncoded systems, the edge pruning versions with $\alpha=0.5\;\text{and}\;1$ do not suffer from performance loss and can even outperform the network without pruning (${\alpha= 0}$). 
This phenomenon is because the pruning operations can remove the redundant connections 
of the FC graph  
and results in a generalizable model less troubled by  overfitting.
Moreover, 
the performance loss in uncoded BER caused by a large proportion of edges being removed when $\alpha$ is high (e.g., $\alpha=2$ and 81.5\% of edges are removed) can be compensated by the strong error correction code. 

{By contrast, the third column of \tabref{tab:impact_ed} shows a clear trade-off between performance and complexity (the proportion of edges reduced) under CCs with a higher code rate, i.e., lower error correction capability.} 
{For example, the edge pruning versions with $\alpha$ equal to 1, 2, and 4 all experience performance loss compared with the original FC model, with the WER increasing from 4.22e-3 at ${\alpha= 0}$ to 7.96e-3 at $\alpha=4$.
However, the computational complexity is significantly reduced since approximately 91.8\% of edges are pruned, and the cost for the message encoding \eqref{eq:propagation} on these edges is saved.}

\CheckRmv{
\begin{figure}[t]
	\centering
	\subfigure[Rayleigh MIMO channel]{
		\includegraphics[width=3in]{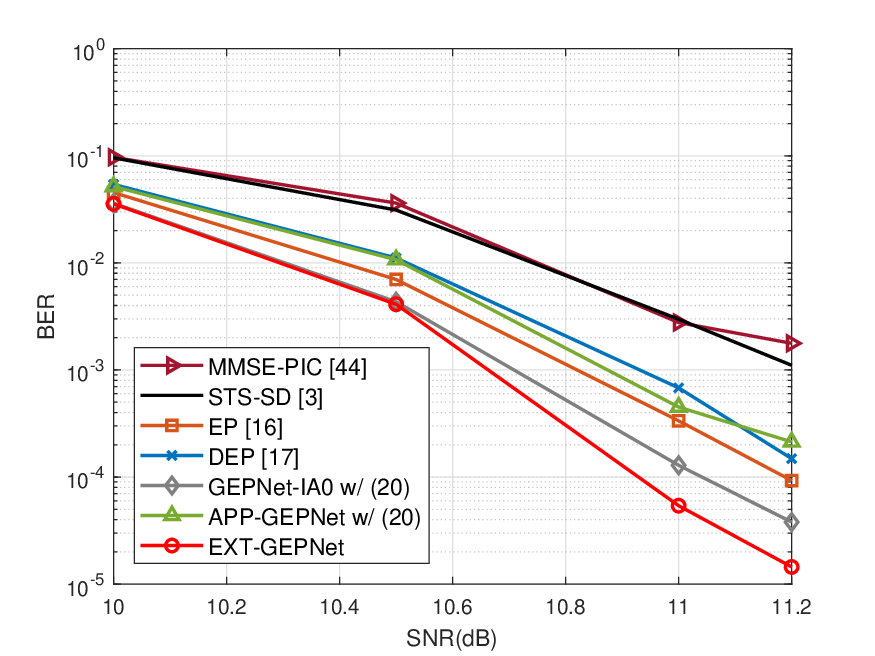}
		\label{fig:ber_4mimo_16qam_pctc}
	}
	\subfigure[Spatially correlated MIMO channel with $\varrho=0.5$]{
		\includegraphics[width=3in]{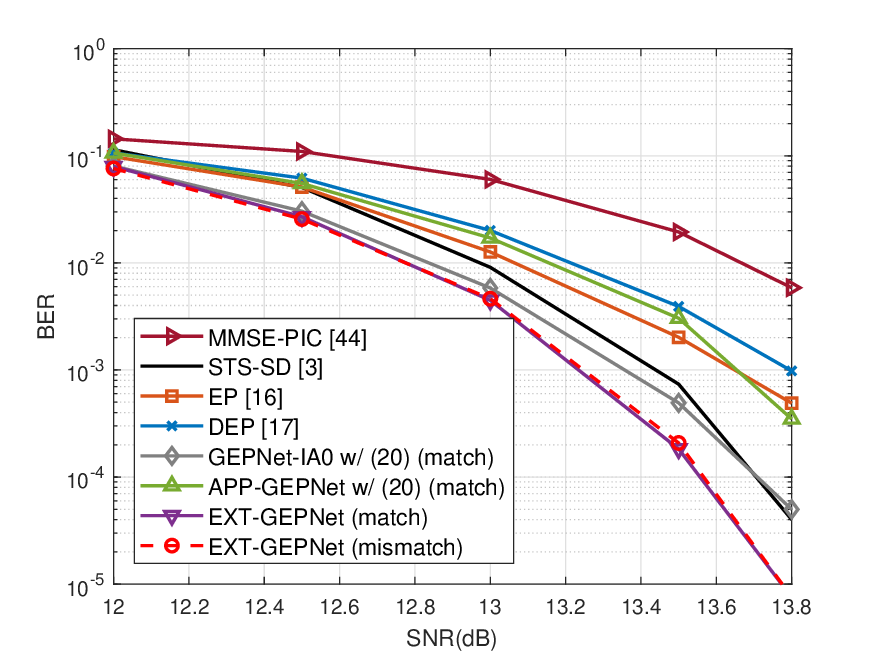}
		\label{fig:ber_4mimo_16qam_corr_pctc}
	}
	\caption{BER performance {($I=2$)} with turbo codes ($R_{\rm c}=1/2,\;N_{\rm b}=1952$) for \Times{4}{4} MIMO channels with 16-QAM.}
	\label{fig:ber_16qam_pctc}
\end{figure}
}

\subsubsection{{Under Turbo Codes and Various Channels}} 
In this subsection, we first demonstrate that the proposed scheme can be applied without dependence on the channel code. 
{\figref{fig:ber_16qam_pctc} provides the BER performance comparison under turbo codes with a code rate of $R_{\rm c}=1/2$ and a word length of $N_{\rm b}=1952$}.
We directly use the network with \SNRTRAIN{=}{13} in \figref{fig:ber_4mimo_16qam_cc12} 
for the comparison under turbo codes and Rayleigh MIMO channel in \figref{fig:ber_4mimo_16qam_pctc} without additional training. 
\figref{fig:ber_4mimo_16qam_pctc} reveals that EXT-{\GEP} generalizes well to the turbo codes and outperforms the other methods {by a large margin}, 
{thereby indicating the advantage of the open-loop training scheme in not relying on the choice of channel codes.}

Furthermore, \figref{fig:ber_4mimo_16qam_corr_pctc} provides the performance comparison under spatially correlated channels to illustrate the robustness of the proposed scheme against channel mismatch.
In particular, the EXT-{\GEP} trained with the Rayleigh channel 
from \figref{fig:ber_4mimo_16qam_cc12} 
is tested under the correlated channel with a spatial correlation coefficient $\varrho=0.5$, and the results are marked by ``mismatch'' in \figref{fig:ber_4mimo_16qam_corr_pctc}.  
The figure demonstrates that the gap between the ``mismatch'' network and the network trained and tested both under the correlated channel can be neglected, thereby verifying the great robustness of the proposed scheme.

\subsubsection{{Robustness to SNR and Antenna Configuration}}
{\figref{fig:ber_4x2mimo_16qam_pctc} illustrates the BER performance of EXT-GEPNet under SNR and antenna configuration mismatches. 
We train an EXT-GEPNet in a \Times{4}{4} MIMO system ($N=K=8$) with \SNR{=}{13} and evaluate the network's performance in a \Times{4}{2} MIMO system ($N=8,K=4$) with varying SNRs.
The channel model is the spatially correlated channel with a correlation coefficient $\varrho=0.5$, which remains consistent during the training and testing phases.
The modulation type used is 16-QAM. 
Turbo codes with  $R_{\rm c}=1/2$ and  $N_{\rm b}=1952$ are applied for channel coding.
The figure reveals that the EXT-GEPNet trained under the mismatched antenna ($N=K=8$) and SNR configurations exhibits the best BER performance among the compared schemes, except for the EXT-GEPNet trained under the matched antenna ($N=8, K=4$) and SNR configurations.
Moreover, the performance gap between the matched and mismatched models is within 0.1 dB, indicating the robustness of the proposed method against antenna configuration and SNR mismatches.}
\CheckRmv{
\begin{figure}[tbp]
	\setlength{\abovecaptionskip}{-0.1cm}
	\setlength{\belowcaptionskip}{-0.0cm}
	\centering
	\includegraphics[width=3in]{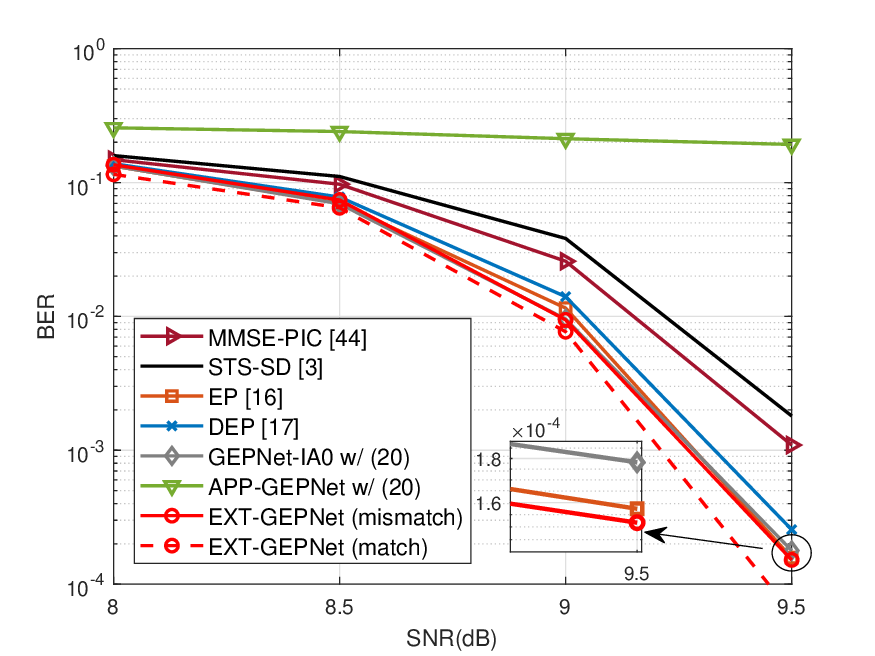}
	\caption{{BER performance ($I=2$) of the EXT-GEPNet with antenna configuration and SNR mismatches for a \Times{4}{2} spatially correlated MIMO channel with $\varrho=0.5$.  Turbo codes with $R_{\rm c}=1/2$ and $N_{\rm b}=1952$ are used. The modulation type is 16-QAM.}}
	\label{fig:ber_4x2mimo_16qam_pctc}
\end{figure}
}

\CheckRmv{
\begin{figure}[tbp]
	\setlength{\abovecaptionskip}{-0.1cm}
	\setlength{\belowcaptionskip}{-0.0cm}
	\centering
	\includegraphics[width=3in]{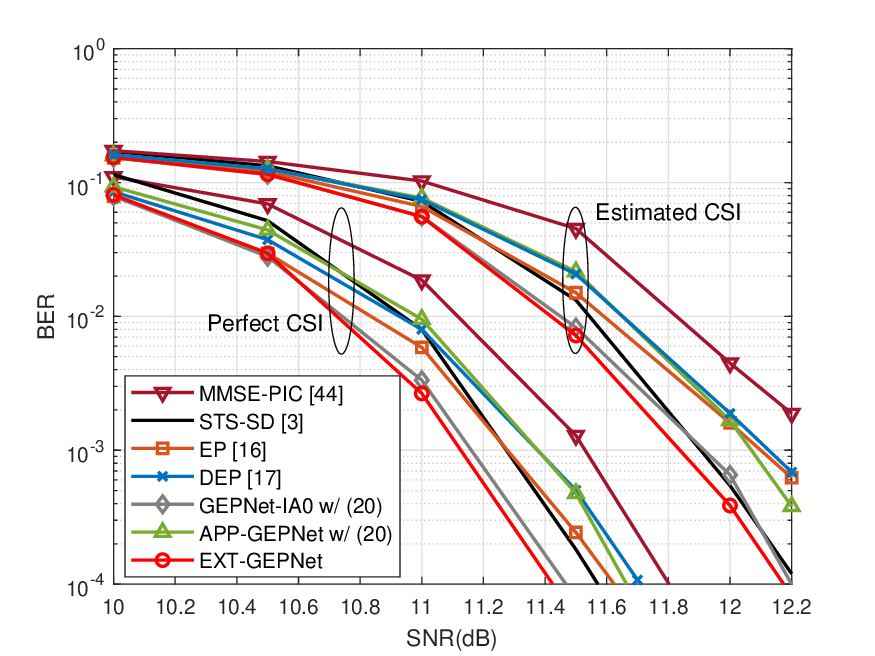}
	\caption{{BER performance comparison ($I=2$) between perfect CSI and estimated CSI using LMMSE method for a \Times{4}{2} correlated MIMO channel specified by the 5G-NR standard. Turbo codes with $R_{\rm c}=1/2$ and $N_{\rm b}=1952$ and 16-QAM are used.}}
	\label{fig:ber_4x2mimo_16qam_nr_pctc}
\end{figure}
}

\subsubsection{{Robustness to Imperfect CSI}}  
{
In the above investigation, all receivers were tested under classical channel models with accurate CSI.
Next, we further validate the proposed method using the spatially correlated channel model from the fifth-generation new radio (5G-NR) \cite{3gpp36101}.
This provides a more realistic evaluation and assesses the algorithm's performance under imperfect CSI to verify its robustness against channel estimation errors.
The considered system contains 2 Tx and 4 Rx antennas as uniform linear arrays.
The channel correlation level is set to medium correlation A, as specified in the 5G-NR standard \cite{3gpp36101}.
We utilize the LMMSE method \cite{FundamentalsStatisticalSignal} with an orthogonal pilot matrix $\mathbf{X}_{\rm p}\in \mathbb{C}^{N_{\rm t}\times N_{\rm p}}$ composed by $N_{\rm t}=2$ columns of the discrete Fourier transform matrix $\mathbf{F}\in \mathbb{C}^{N_{\rm p}\times N_{\rm p}}$ to estimate the complex-valued channel matrix of size $4 \times 2$, where $N_{\rm p}$ is the number of pilot vectors in a time slot and is set to 16. 
Turbo codes with $R_{\rm c}=1/2$ and $N_{\rm b}=1952$ and 16-QAM modulation are used.
\figref{fig:ber_4x2mimo_16qam_nr_pctc} illustrates the BER performance under perfect and estimated CSIs. 
The neural networks are trained using the 5G-NR channel model and \SNRTRAIN{=}{10}. 
As expected, all the tested algorithms experience a similar performance loss when transitioning from perfect to imperfect CSIs, indicating that the accuracy of CSI significantly influences the detection accuracy.
However, EXT-GEPNet consistently outperforms the other tested algorithms, demonstrating its robustness to imperfect CSI. This observation holds true regardless of the presence of channel estimation errors, thereby highlighting the effectiveness and resilience of the proposed algorithm.}

\newcolumntype{M}[1]{>{\arraybackslash}m{#1}}
\CheckRmv
{
	\begin{table*}
	\centering
	\begin{threeparttable}
	\caption{Computational complexity of different turbo receivers}
	\begin{tabular}{l|l|l|c|c} 
	\toprule
	\multicolumn{2}{l|}{\multirow{2}{*}{Algorithms}}   & \multicolumn{1}{c|}{\multirow{2}{*}{Computational complexity: $C_{\text{det}1}+(I-1)C_{\text{det}\iota}$}} & \multicolumn{2}{c}{Example$^{\mathrm a}$}   \\ 
	\cline{4-5}
	\multicolumn{2}{l|}{}  &    & RVMs  & Time  \\ 
	\midrule
	\multicolumn{2}{l|}{\multirow{2}{*}{MMSE-PIC \cite{witzkeIterativeDetectionMIMO2002}}}   & $C_{\MMSEPIC 1}=N K^{2}+N K+K^{3}+4 K^{2}+(M+3) K$  & \multirow{2}{*}{$2.91\times 10^3$}  & \multirow{2}{*}{$3.39\times 10^{-4}$} \\
	\multicolumn{2}{l|}{} & $C_{\MMSEPIC \iota}=C_{\MMSEPIC 1}+3MK$  &  &   \\ 
	\hline
	\multicolumn{2}{l|}{\multirow{2}{*}{EP{\cite{santosSelfTurboIterations2019}}}} & $C_{\EP 1}=N K^{2}+N K+(K^{3}+K^{2}+13 K+2 M K) T$   & {\multirow{2}{*}{$9.01\times 10^3$}} & \multirow{2}{*}{$3.52\times 10^{-4}$}  \\
	\multicolumn{2}{l|}{}  & {$C_{\EP \iota}=C_{\EP 1} + 2MKT+ 3MK$}  &  &  \\ 
	\hline
	\multicolumn{2}{l|}{\multirow{2}{*}{{DEP \cite{murillo-fuentesLowComplexityDoubleEPBased2021}}}} & {$C_{\DEP 1}=N K^{2}+N K+(K^{3}+K^{2}+13 K+2 M K) T$}   & {\multirow{2}{*}{$9.17\times 10^3$}} & {\multirow{2}{*}{$3.71\times 10^{-4}$}} \\
	\multicolumn{2}{l|}{}  & {$C_{\DEP \iota}=C_{\DEP 1} + 2MKT+ 8MK$}  &  &  \\ 
	\hline
	\multicolumn{2}{l|}{STS-SD \cite{studerSoftInputSoft2010}} & \multicolumn{1}{c|}{\diagbox { }{ }}  & \diagbox { }{ }  & $1.68\times 10^{-2}$  \\ 
	\hline
	EXT-GEPNet & \begin{tabular}[c]{@{}l@{}}${\alpha= 0}, \eta=1$ \\$\alpha=0.5, \eta=0.410$ \\$\alpha=1, \eta=0.313$ \\$\alpha=2, \eta=0.186$\\$\alpha=4, \eta=0.066$\end{tabular} & \begin{tabular}[c]{@{}M{25em}@{}}$C_{\GEPNet 1}= \big((2N_{\rm u}+2)N_{\rm h1} + N_{\rm h1}N_{\rm h2} + N_{\rm h2}N_{\rm u}\big)LT K(K-1)\cdot\eta + \big(4N_{\rm u} + 3N_{\rm h1} + 9\big)N_{\rm h1}KLT+\big(K^3 + K^2+13K + 2MK + (N_{\rm u}N_{\rm h1}+N_{\rm h1}N_{\rm h2}+N_{\rm h2}M)K\big)T $\\$C_{\GEPNet \iota}=C_{\GEPNet 1} +  K\log_2M +2MKT+3MK$\end{tabular}                                     
	& \begin{tabular}[c]{@{}c@{}}{$6.48\times 10^6$}\\$4.20\times 10^6$\\$3.82\times 10^6$\\$3.33\times 10^6$\\$2.87\times 10^6$\end{tabular} & \multicolumn{1}{c}{\begin{tabular}[c]{@{}c@{}}$6.64\times 10^{-3}$\\$5.74\times 10^{-3}$\\$5.21\times 10^{-3}$\\$4.71\times 10^{-3}$\\$4.21\times 10^{-3}$\end{tabular}}  \\
	\bottomrule
	\end{tabular}
	\label{tab:complexity}
	\begin{tablenotes}[para,flushleft]
		\footnotesize
		\item[]$^{\mathrm{a}}$ The average number of RVMs and running time (in Seconds) 
		% (averaged over SNR range of [10,15] dB for STS-SD) 
		for the detection of one symbol vector when $N=K=8, M=4,N_{\rm h1}=64,N_{\rm h2}=32,N_{\rm u}=8,L=2,T=5,$ and $I=2$. All the algorithms are implemented on the same PC with an Intel Core i7-11700 CPU @ 2.50 GHz and 16 GB memory.
	\end{tablenotes}
	\end{threeparttable}
	\end{table*}
}

\subsection{Computational Complexity Analysis} \label{sec:complexity}

% ok
In this subsection, we analyze the computational complexity of different turbo receivers, focusing on MIMO detection complexity since the channel decoding part for different receivers is the same. 
\tabref{tab:complexity} provides the number of real-valued multiplications (RVMs) and running time required for detecting one symbol vector. 
% logic: ok
The per symbol vector detection complexity is $C_{\text{det}1}+(I-1)C_{\text{det}\iota}$, where $C_{\text{det}1}$ denotes the complexity of the first {\TI} without utilizing \prior information, and {$C_{\text{det}\iota}$} is the counterpart with the priors considered.
% ext-gep ok

The $C_{\text{det}1}$ of EXT-{\GEP}, denoted as $C_{\GEPNet 1}$, can be divided into the operations for GNN and EP in each layer of the network. 
%% gnn-- one shot ok
The complexity of the GNN operations contains the cost for the propagation, aggregation, and readout steps. 
The propagation step is involved in the $L$-round of MP, and each round costs $\big((2N_{\rm u}+2)N_{\rm h1} + N_{\rm h1}N_{\rm h2} + N_{\rm h2}N_{\rm u}\big) K(K-1)$ RVMs for the execution of the function $\mathcal{M}$ in \eqref{eq:propagation} by $K(K-1)$ times in the FC model. % L round of MP
We further consider the effect of edge pruning in the complexity expression listed in the table, where the cost for this step is reduced by the scale $\eta$, which is the proportion of the remaining edges after pruning.
In addition, the aggregation step performs $L$ times, and the readout step is conducted once for one forward inference of the GNN, resulting in a cost of $\big(4N_{\rm u} + 3N_{\rm h1} + 9\big)N_{\rm h1}KL$ and $(N_{\rm u}N_{\rm h1}+N_{\rm h1}N_{\rm h2}+N_{\rm h2}M)K$ RVMs, respectively.  
%% ep ok 
The complexity of the EP procedure is dominated by the matrix inversion in \eqref{eq:sigma}, which requires $K^3+K^2+2K$ RVMs. 
Meanwhile, the matrix-vector multiplications involved in \eqref{eq:cavity}, \eqref{eq:ep_post}, \eqref{eq:post_ext}, and \eqref{eq:damping} cost another $2MK+11K$ RVMs \cite{kosasih2022graph}.

{Additionally, the number of layers $T$ should be considered for $C_{\text{det}1}$ of EXT-{\GEP}, EP, and DEP, as shown in \tabref{tab:complexity}, because all of them are iterative methods.}  
{The additional operations for $C_{\text{det}\iota}$ include the incorporation of non-uniform priors for computations of the initial pair and estimated APPs \cite{santosSelfTurboIterations2019}.} 
The corresponding results are listed in \tabref{tab:complexity}.

% ok
Subsequently, we present a numerical demonstration of the computational complexity for the \Times{4}{4} MIMO system, which is considered in the simulation of the turbo receiver. 
The number of RVMs of the competing schemes in this system is presented in the last but one column of \tabref{tab:complexity}. 
% ok 
EXT-{\GEP} requires more RVMs than EP because of the additional GNN operations.
Notably, the GNN operations can be more computationally expensive than the matrix inversion in EP for the considered small-sized \Times{4}{4} system because the chosen hyperparameters of the GNN ($N_{\rm h1}=64$ and $N_{\rm h2}=32$) are much larger than the system sizes ($N=K=8$).  
% ok
However, the matrix inversion becomes dominant when the system size $K$ grows, because the complexity of matrix inversion is on the order of $\mathcal{O}(K^3)$, whereas the complexity of GNN is on the order of $\mathcal{O}(K^2)$, as indicated in \tabref{tab:complexity}.
Therefore, the complexity ratio between the proposed EXT-{\GEP} and EP in large-scale MIMO systems narrows down and tends to 1. 

% ok
Additionally, we analyze the effect of edge pruning on the numerical complexity results.
The pruning factor $\alpha$ of EXT-{\GEP} for calculating the number of RVMs is with the same choice as that for the performance evaluation in \tabref{tab:impact_ed}. 
The column of RVMs in \tabref{tab:complexity} shows that a larger $\alpha$ leads to a smaller $\eta$ and thus fewer RVMs, as expected.  
% ok
For example, approximately 35\%, 41\%, and 49\% RVMs can be saved using a pruning factor $\alpha$ equal to 0.5, 1, and 2, respectively. 
Considering the performance under $R_{\rm c}=1/2$ given by {the second column of} \tabref{tab:impact_ed},
the edge pruning versions with $\alpha$ equal to 0.5, 1, and 2 reach a more desirable trade-off between performance and complexity than the original EXT-{\GEP} without pruning.

Finally, we compare the complexity of EXT-{\GEP} with STS-SD \cite{studerSoftInputSoft2010}.
The number of RVMs for STS-SD  is omitted in \tabref{tab:complexity}, as there is no analytical complexity expression, and the required operations are mainly sequential search \cite{studerSoftInputSoft2010}, which cannot be evaluated by multiplication count. 
% ok
Therefore, we perform a running time test for the comparison under
the \Times{4}{4} system configuration.  
Results reveal that the proposed EXT-{\GEP} without pruning requires 2.53 times less running time than STS-SD.
The improvement of the edge pruning versions in reducing the running time is also remarkable.
For example, the edge pruning version with $\alpha=4$ runs 1.58 and 3.99 times faster than the FC EXT-{\GEP} and STS-SD, respectively.

Moreover, the STS-SD is impractical to realize for large-scale MIMO systems because its complexity is exponential to the system size \cite{jaldenComplexitySphereDecoding2005}.  
In contrast to STS-SD, the proposed scheme has a complexity that polynomially grows with the system sizes and can make a flexible trade-off between complexity and performance by adjusting the pruning factor $\alpha$, as well as the hidden layer sizes $N_{\rm h1}$ and $N_{\rm h2}$ in the GNN.  
Therefore, EXT-{\GEP} can be viewed as a powerful and efficient scheme given the performance indicated in the simulations and the polynomial complexity reported in \tabref{tab:complexity}.

% \vspace{-0.5cm}
%%%%%%%%%%%%%%%%%%%%%%%%%%%%%%%%%%%%%%%%%%%%%%%%%%%%%%%%%%
\section{Conclusions} \label{sec:conclusion} 
{We proposed a GNN-enhanced EP algorithm for MIMO turbo receivers.
We first developed the soft-input soft-output mechanism for {\GEP} and the corresponding turbo receiver structure.
We then customized a training scheme to establish the EXT-{\GEP},  
which inherits the superiority of {\GEP} in achieving an augmented \post estimation via the GNN and addresses the limitations of failing to produce reliable extrinsic LLRs.
The EXT-{\GEP} can be deployed in the developed turbo structure to take full advantage of various prior information and achieve stable turbo receiving, outperforming the approach using the original {\GEP}.}
Furthermore, we developed an edge pruning method to eliminate the redundancy in the network, 
{resulting in a significant complexity reduction with negligible performance loss.}
{Complexity analysis and simulation results confirm the efficiency, excellent performance, and adaptability of the proposed scheme.}

%%%%%%%%%%%%%%%%%%%%%%%%%%%%%%%%%%%%%%%%%%%%%%%%%%%%%%%%%%
% \section*{Acknowledgment}

% The authors would like to thank

% Can use something like this to put references on a page
% by themselves when using endfloat and the captionsoff option.
\ifCLASSOPTIONcaptionsoff
  \newpage
\fi

% \bibliographystyle{IEEEtran}      
% \bibliography{IEEEabrv,ref}

% Generated by IEEEtran.bst, version: 1.14 (2015/08/26)

\end{document}